\documentclass{aa}  

\usepackage{graphicx}
\usepackage{txfonts}
\usepackage{lipsum}
\usepackage{subcaption}   
\usepackage{caption}

\usepackage{lscape}  
\usepackage{amsmath}

\usepackage{placeins}
\usepackage{xcolor}

\usepackage[hidelinks]{hyperref}

\begin{document}

   \title{Coronal gas excitation as a tracer of supermassive black hole mass: on the mid-IR coronal [\ion{Ne}{v}] lines}

   \author{J. S. Elford\inst{1,2}\fnmsep\thanks{Corresponding author: jacob.elford@iac.es}, A. Prieto\inst{1,2}, A. Rodríguez-Ardila{\inst{3}}
        }

   \institute{$^1$ Instituto de Astrofísica de Canarias, Calle Vía Láctea, s/n, E-38205 La Laguna, Tenerife, Spain\\
   $^2$ Departamento de Astrofísica, Universidad de La Laguna, E-38206 La Laguna, Tenerife, Spain\\
   $^3$ Observatório Nacional, Rua General José Cristino 77, CEP 20921-400, São Cristóvão, Rio de Janeiro, RJ, Brazil}

   \date{Received September 30, 20XX}

  \abstract
   {}
   {Coronal lines (CL) may be a reliable proxy of supermassive black hole (SMBH) masses in active galactic nuclei (AGN) because the production of these high ionisation potential (IP) lines is sensitive to the shape of the ionising continuum which, for a thin accretion disc, relies on the black hole mass. In this work we study the connection between the [\ion{Ne}{v}] coronal lines and the SMBH mass.} 
   {We analyse the \textit{Spitzer spectra} of a sample of 27 AGN spanning three orders of magnitude in SMBH mass. Fluxes and the electron density in the coronal gas medium were measured after fitting Gaussians to both the $\rm [\ion{Ne}{v}]14\mu m$ and $\rm [\ion{Ne}{v}]24\mu m$ lines where detected at high signal to noise.}
   {Line fluxes were normalised to the Br$\gamma_{\rm broad}$ emission from the broad line region. We found strong correlations between the $\rm [\ion{Ne}{v}]14\mu m/Br\gamma_{\rm broad}$ and $\rm [\ion{Ne}{v}]24\mu m/Br\gamma_{\rm broad}$ line ratios and the SMBH mass. For this sample of 27 AGN we find scatters of 0.54 and 0.53 dex, respectively. These correlations support the theory that the coronal gas excitation is sensitive to a range of SMBH masses because of the dependence on the shape of the ionising continuum. The electron densities of the coronal medium are of the order 1000$\rm cm^{-3}$ supporting their nuclear origin. A comparison with density values derived from JWST for a subsample of objects confirm this conclusion.}
   {These results further demonstrate the ability to use CL as a SMBH mass proxy, allowing for accurate SMBH mass measurements in AGN.}
   \keywords{accretion, accretion discs – radiation mechanism thermal – techniques: spectroscopic – galaxies: active – quasars:
emission lines}
    \titlerunning{mid-IR [\ion{Ne}{v}] lines as a tracer of SMBH mass}
   \maketitle
    \nolinenumbers

\section{Introduction}
Tight correlations have been found between supermassive black hole (SMBH) masses and the properties of host galaxies (such as bulge mass \citealp[e.g.][]{Magorrian1998_M_Bulge,MarconiHunt2003_M_Bulge} and velocity dispersion \citealp[e.g.][]{FerraresMerritt2000_M_Sigma,Tremaine2002_M_Sigma,Gultekin2009_M_Sigma}). These tight correlations could suggest a co-evolution between SMBHs and their host galaxies \citep[see e.g.][for a review]{KormendyHo2013_SMBH_Co_Evo}. However, to study and understand this co-evolution accurate measurements of the SMBH masses are needed. There are several methods to measure the masses of SMBH, including modeling the stellar \citep[e.g.][]{Cappellari2002_BH_Mass_Stellar,Krajnoic2009_BH_Mass_Stellar}, ionized gas \citep[e.g.][]{Ferrarese1996_BH_Mass_Ionized,Sarzi2001_BH_Mass_Ionized,DallaBonta2009_BH_Mass_Ionized,Walsh2013_BH_Mass_Ionized}, maser \citep[e.g.][]{Miyoshi1995_BH_Mass_MASER,Greene2010_BH_Mass_MASER,Kuo2011_BH_Mass_MASER} and molecular gas kinematics \citep[e.g.][]{Davis2017_BH_Mass_Molecular,Onishi2017_BH_Mass_Molecular,North2019_BH_Mass_Molecular,Smith2019_BH_Mass_Molecular,Davis2020_BH_Mass_Molecular,Smith2021_BH_Mass_Molecular,Lelli2022_BH_Mass_Molecular,Ruffa2023_BH_Mass_Molecular,Zhang2024_BH_Mass_Molecular,Dominiak2025_BH_Mass_Molecular,Zhang2025_BH_Mass_MolecularB,Zhang2025_BH_Mass_MolecularA}. Other methods rely on the correlations between the SMBH mass and host galaxy properties (such as the bulge mass and the velocity dispersion) to infer the SMBH mass. However, these methods can be observationally expensive, may only be applicable to certain sources or have large scatters in the derived relationship. Additionally, in active galactic nuclei (AGN) using these methods to measure the black hole mass is further complicated. The energetic output can disturb the gas surrounding it, making modeling difficult or the strong continuum emission from the AGN can dilute the stellar absorption lines, making measurements of the velocity dispersion harder to determine.\par
There have been efforts to find a scaling relation that allows us to measure the SMBH masses in AGN. It has been found that the AGN continuum luminosity (optical, UV or X-ray) correlates with the size of the AGN Broad Line Region (BLR), which in turn can be used to estimate the SMBH mass known as the mass-luminosity relation \citep[e.g.][]{Koratkar1991_Mass_Luminosity,Kaspi2000_Mass_Luminosity,Kaspi2005_Mass_Luminosity,Landt2013_Mass_Luminosity}. However, the mass-luminosity relation can have a scatter of 40\% \citep{Kaspi2005_Mass_Luminosity} driven by changes in the optical-UV continuum shape with increasing AGN luminosity. Other scaling relations have been proposed using emission lines such as [\ion{O}{iii}]\,$\lambda5007$ to measure the velocity dispersion \citep[e.g.][]{NelsonWhittle1996_OIII_Bulge}, [\ion{O}{ii}]\,$\lambda3727$, H\,$\beta$ or H\,$\alpha$ \citep[e.g.][]{Kaspi2005_Mass_Luminosity,GreeneHo2005_BH_Mass_Halpha} to infer the size of the BLR or [\ion{Fe}{ii}] to measure the velocity dispersion \citep[e.g.][]{Riffel2013_FeII_Sigma}. However, these relations can also have very large scatters giving large uncertainties on the SMBH masses. The use of single epoch broad line estimates is widely used to estimate SMBH masses in AGN. This method has been largely applied, for example, to data from the Sloan Digital Sky Survey (SDSS) in a large number of AGN \citep[e.g.][]{DallaBonta2020_SDSS_AGN_BH_Mass,DallaBonta2025_SDSS_AGN_BH_Mass}. However, other studies have found this method could overestimate the SMBH masses by an average of $~$0.3 dex \citep[e.g.][]{FonsecaAlvarez2020_BH_Mass_Error}. This motivates the search for new approaches to robustly estimate SMBH masses in AGN that can be applied to a wide sample of sources.\par
With the James Webb Space Telescope (JWST) and Vera Rubin Observatory's LSST starting operation, the search for SMBH mass scaling relations in the near to mid infrared (IR) regime is ongoing. Specifically, using coronal lines (CL) might give rise to a novel and powerful scaling relation to accurately measure the SMBH mass in AGN. CL can be found in the X-ray, optical and IR regimes and have high ionisation potentials (IP) from $\sim$100\,eV to a few hundreds eV which make them an excellent tracer of the AGN ionising continuum. Although often fainter than the classical emission lines used in diagnostic diagrams, high angular resolution observations of nearby AGN in the near-IR have found that CL are very prominent features \citep[e.g.][]{Marconi1994_Coronal_Lines,Reunanen2003_Coronal_Lines,Prieto2005_Coronal_Lines,Rodriguez-Ardila2006_Coronal_Lines,Muller-Sanchez2011_Coronal_Lines,Rodriguez-Ardila2017_Coronal_Lines,Gravity2020_Coronal_Lines}.\par
The use of CL as a black hole mass proxy was proposed by \cite{Cann2018_CL_Theory}, which showed via photoionisation simulations that in the ($10^2-10^5\,\rm M_\odot$) mass range that high IP CL are produced more favourably than CL with a lower IP. Following from this \cite{Prieto2022_SiVI_BH} found a correlation between the SMBH mass in the $10^6-10^8\,\rm M_\odot$ mass range and the [\ion{Si}{vi}]\,$1.963\mu m$/Br$\gamma_{\rm broad}$ line ratio. They showed, using photoionisation modeling that this correlation is likely driven by the accretion disc temperature in this mass range, allowing for the optimal production of the [\ion{Si}{vi}] CL, which has an IP=167\,eV. Other CLs studied in that work with higher IP did not correlate with the SMBH mass, possibly because the disc temperature is lower than that implied by the high IP. These studies together suggest the CL that are most sensitive to the SMBH mass shifts towards lower energies as the accretion disc temperature decreases with increasing SMBH mass.\par
In this paper we follow up the work done in \cite{Prieto2022_SiVI_BH} by studying a similar sample of sources but focusing on the [\ion{Ne}{v}]$\rm 14\mu m$ and $\rm24\mu m$ lines, which both have a slightly lower IP than [\ion{Si}{vi}] (IP=97\,eV for [\ion{Ne}{v}] vs 1P=168\,eV for [\ion{Si}{vi}]). If a correlation is found with both of these [\ion{Ne}{v}] CL, it will be consistent with the interpretation that a correlation between CL and SMBH masses is driven by the disc temperature allowing for optimal production of these CL and will further strengthen the use of CL as a SMBH mass proxy. The [\ion{Ne}{v}] line is ideal for this study as it is bright in the MIR and is covered by both \textit{Spitzer} and JWST, giving us an abundance of archival data to work with and allowing for this study to be extended with future observations. We focus on the $\rm [\ion{Ne}{v}]/Br\gamma_{broad}$ ratio because $\rm Br\gamma_{broad}$ gives a measure of the overall ionising output from the AGN whereas the [\ion{Ne}{v}] CL traces the hard, high energy part of the continuum. By using the ratio of these two, it allows us to directly probe the shape of the ionising continuum. The shape of the ionising continuum is set by the temperature of the AGN accretion disc, which in turn depends on the SMBH mass, so a relation between this ratio and the SMBH is physically motivated. We focus on the $\rm Br\gamma$, as this line is a bright hydrogen recombination line in the NIR that allows us to probe the total ionising output of the AGN. We select it over other hydrogen recombination lines, as it is close in wavelength to the CL studied, reducing the potential impact of extinction. We focus on the broad component of this line, as it is produced directly by the AGN and is more likely to be coupled to the production of the CL. In contrast, the narrow component originates on a much larger scale, which could add additional scatter or even completely wash out any potential relations.\par
The paper is organised as follows. In Section \ref{Sample} we introduce the sample studied in this work and explain the data collected. In Section \ref{Results} we present the results of this work, which we then discuss in Section \ref{Discussion}. We finally summarise and conclude the paper in Section \ref{Conclusion}. In this paper we adopt the following cosmology: $\rm H_0=70\,km\,s^{-1}\,Mpc^{-1}$, $\rm \Omega_m=0.30$, $\Omega_{\Lambda}=0.70$.

\section{Sample and Data}\label{Sample}
The sample was selected from \textit{The AGN Black Hole Mass Database} \citep{Bentz2015_AGN_BH_Mass_Database} which is a collection of 86 nearby AGN with SMBH masses estimated by reverberation mapping. This sample is exclusively Type 1 AGN, as the broad lines are the main focus of reverberation mapping campaigns. This sample contains a range of Type 1 AGN classifications (e.g. Seyfert 1.8, narrow-line Seyfert 1, radio-loud AGN, etc.). Due to the high signal-to-noise required for reverberation mapping, this does bias us towards brighter sources. In future work, we will expand our sample to include low-luminosity AGN (LLAGN). From this sample of 86 sources, we then select sources with archival \textit{Spitzer} data and published Br$\gamma_{\rm broad}$ measurement. This leaves us with a final sample of 27 sources. No selection criteria was employed regarding Type 1 activity or AGN luminosity. The final sample of sources overlaps strongly with those studied in \cite{Prieto2022_SiVI_BH}.\par
The BH masses were mostly taken from \cite{Bentz2015_AGN_BH_Mass_Database} with coronal line ratios either measured in spectra presented in this work or published values. The full sample of sources is shown in Table \ref{tab:sample}. 
\subsection{MIR Spectroscopy}
To measure the $[\rm \ion{Ne}{v}]\,14\mu m$ and $[\rm \ion{Ne}{v}]\,24\mu m$ line fluxes we retrieved the spectra from the Cornell Atlas of \textit{Spitzer}/\textit{IRS} Sources (CASSIS) archive \citep{Lebouteiller2011_CASSIS}. We only used the high-resolution observations for each source with the $[\rm \ion{Ne}{v}]\,14\mu m$ line falling in the Short-High (SH) 13 module and the $[\rm \ion{Ne}{v}]\,24\mu m$ falling in the Long-High (LH) 16 module. The high-resolution IRS modules have a spectral resolution (R) $\approx600$. The aperture sizes (\farcs) of the SH and the LH modules are $4.7\times11.3$ and $11.1\times22.3$, respectively. We use the optimally extracted one-dimensional spectra from the CASSIS archive with no additional spectral extraction performed. To measure the line flux, we first defined a fitting region of 0.2$\rm \mu m$ from either side of the centroid position of the line. We then masked out the line assuming the line is in the central 0.2$\mu m$ of the fitting region. Following from this masking procedure, we then fit the continuum assuming a flat continuum shape which, after visual inspection of the spectrum, accurately reproduces the local continuum shape. We then subtract the estimated continuum from the spectrum. Using this continuum-subtracted spectra, we then fit a single Gaussian to the line with the amplitude (A), the line center ($\mu$) and the line width ($\sigma$) as free parameters. The fits were then visually inspected to ensure high-quality fits of the line and to ensure no nearby features were impacting the fitting procedure. After this visual inspection we found these single Gaussians accurately reproduced the shape of the NeV lines and no nearby features were impacting the fits. We used the errors provided with the CASSIS one-dimensional \textit{Spitzer} spectra, which is a combination of the RMS and systematic errors, and treated them as absolute in our fitting. When estimating the errors in our measured line fluxes, we propagate the errors in the amplitude and the line width in quadrature. To measure the S/N, we use the measured line fluxes and the uncertainty on these line fluxes. If a fit had a S/N less than 3 we did not include it in our analysis. The remaining fits are able to reproduce the shape of the [\ion{Ne}{v}] lines with S/N ranging from 3 to 41. The \textit{Spitzer} spectra of both [\ion{Ne}{v}] lines and the associated Gaussian fits are shown in Figure \ref{Spitzer_Fits}. In the case of 3C273 the $[\rm \ion{Ne}{v}]\,14\mu m$ was not well fitted due to low S/N, so we took the published value from \cite{Fernandez-Ontiveros2016_CL_Density}. In the case of MRK707, neither line was well fitted again due to low S/N so we instead take the $[\rm \ion{Ne}{v}]\,14\mu m$ line flux from \cite{Spoon2022_MRK707_NeV}. In the situations where multiple \textit{Spitzer} spectra are available for a source, we take a weighted average of the measured lines. 
\subsection{NIR Spectroscopy}
We retrieve the Br$\gamma_{\rm broad}$ from \cite{Prieto2022_SiVI_BH}. The majority of these fluxes were extracted from \cite{Riffel2006_Line_Ratios}. If they were not available in this work, observations were obtained with the Gemini Near-Infrared Spectrograph (GNIRS) attached to the Gemini North Telescope or the ARCOiRIS spectrograph, mounted on either the Blanco or SOAR Telescopes. See \cite{Prieto2022_SiVI_BH} for full details on these observations and the data reduction procedures. For six sources we instead took the Br$\gamma_{\rm broad}$ measurements from \cite{Lamperti2017_BAT_BrG}. These measurements were obtained from an independent dataset and analysis than those in \cite{Riffel2006_Line_Ratios}. Differences in flux calibration, decomposition methodology and intrinsic AGN variability may introduce additional scatter, but no evidence was found that these effects dominate the observed correlations, and no correction was applied. The decomposition into broad and narrow components for both these works was performed by automated spectral fitting programs. We refer the reader to those works and the references within for full details on this decomposition procedure. Differences in the adopted decomposition methodology may contribute to the scatter in these relations.
\begin{table*}[]
    \centering
    \caption{Sample used in this work}
    \begin{tabular}{ccccccccc}
    \hline \\
    Galaxy & ${\rm log}M_{\rm BH}$ & $[\rm \ion{Ne}{v}]\,14\mu m$ & $[\rm \ion{Ne}{v}]\,24\mu m$ & Br$\gamma_{\rm broad}$ & ${\rm log}\frac{[\rm \ion{Ne}{v}]}{\rm Br\gamma_{\rm broad}}$ & ${\rm log}\frac{[\rm \ion{Ne}{v}]\,24\mu m}{\rm Br\gamma_{\rm broad}}$ & $\frac{[\rm \ion{Ne}{v}]\,24\mu m}{[\rm \ion{Ne}{v}]\,14\mu m}$ & $\rm log(n_e)$\\
    (1) & (2) & (3) & (4) & (5) & (6) & (7) & (8) & (9)\\
    \hline
    3C120 & $7.74\pm0.04$ & $13.6\pm0.43$ & $21.0\pm0.65$ & $4.63\pm0.66$ & $0.47\pm0.06$ & $0.66\pm0.06$ & 1.54 & <3\\
    3C273 & $8.84\pm0.10$ & $2.12\pm0.42^*$ & $2.52\pm0.27$ & $4.22\pm0.66$ & $-0.30\pm0.11$ & $-0.22\pm0.08$ & 1.19 & 3\\
    \textit{ARK120} & $8.07\pm0.06$ & $1.11\pm0.23$ & $1.63\pm0.44$ & $7.24\pm0.18$ & $-0.81\pm0.09$ & $-0.65\pm0.12$ & $1.46$ &<3\\
    ARK564 & $6.59\pm0.17$ & $7.34\pm0.29$ & $6.47\pm0.36$ & $0.57\pm0.04$ & $1.11\pm0.03$ & $1.05\pm0.04$ & 0.88 & 3\\
    Fairall9 & $8.29\pm0.09$ & $3.10\pm0.27$ & $2.50\pm0.16$ & $4.31\pm0.47$ & $-0.14\pm0.06$ & $-0.24\pm0.06$ & 0.81 & 3.5\\
    \textit{MRK79} & $7.61\pm0.12$ & $9.53\pm0.44$ & $11.6\pm0.35$ & $2.58\pm0.07$ & $0.57\pm0.02$ & $0.65\pm0.02$ & 1.22 & 3\\ 
    \textit{MRK279} & $7.44\pm0.12$ & $3.24\pm0.24$ & $2.03\pm0.13$ & $1.12\pm0.03$ & $0.46\pm0.03$ & $0.26\pm0.03$ & $0.63$ & 3.5\\
    MRK335 & $7.23\pm0.04$ & $1.08\pm0.13$ & $1.31\pm0.12$ & $2.67\pm0.31$ & $-0.40\pm0.07$ & $-0.31\pm0.06$ & 1.23 & 3\\
    MRK509 & $8.05\pm0.04$ & $6.24\pm0.21$ & $7.46\pm0.22$ & $34.9\pm2.18$ & $-0.75\pm0.03$ & $-0.67\pm0.03$ & 1.20 & 3\\
    MRK707 & $6.50\pm0.15$ & $0.83\pm0.09^*$ & - & $0.48\pm0.04$ & $0.24\pm0.06$ & - & - & -\\
    MRK766 & $6.82\pm0.05$ & $21.9\pm0.47$ & $18.8\pm0.47$ & $2.00\pm0.23$ & $1.04\pm0.05$ & $0.97\pm0.05$ & 0.86 & 3.5\\
    \textit{MRK817} & $7.59\pm0.07$ & $1.82\pm0.30$ & - & $2.05\pm0.06$ & $-0.05\pm0.07$ & - & - & -\\
    MRK841 & $8.10\pm0.02$ & $8.59\pm0.23$ & $5.48\pm0.58$ & $2.61\pm0.75$ & $0.52\pm0.13$ & $0.32\pm0.13$ & 0.64 & 4\\
    NGC863 & $7.57\pm0.06$ & $0.75\pm0.20$ & - & $1.04\pm0.29$ & $-0.14\pm0.17$ & - & - & -\\
    NGC3227 & $6.78\pm0.10$ & $23.4\pm0.78$ & $16.5\pm0.50$ & $2.00\pm0.39$ & $1.07\pm0.09$ & $0.92\pm0.09$ & 0.71 & 3.5\\
    \textit{NGC3516} & $7.40\pm0.05$ & $4.97\pm0.42$ & $7.88\pm0.47$ & $4.70\pm1.32$ & $0.02\pm0.13$ & $0.22\pm0.12$ & 1.59 & <3\\
    NGC3783 & $7.37\pm0.06$ & $16.8\pm0.60$ & $13.2\pm0.91$ & $6.04\pm0.98$ & $0.44\pm0.07$ & $0.34\pm0.08$ & 0.79 & 3.5\\
    NGC4051 & $6.13\pm0.14$ & $11.6\pm0.34$ & $9.05\pm0.28$ & $1.31\pm0.08$ & $0.95\pm0.03$ & $0.84\pm0.03$ & 0.78 & 3.5\\
    NGC4151 & $7.55\pm0.05$ & $77.3\pm1.84$ & $55.1\pm4.03$ & $12.5\pm1.08$ & $0.79\pm0.04$ & $0.64\pm0.05$ & 0.71 & 3.5\\
    NGC4395 & $5.45\pm0.13$ & $1.33\pm0.12$ & $1.36\pm0.19$ & $0.19\pm0.02$ & $0.84\pm0.06$ & $0.85\pm0.07$ & 1.02 & 3\\
    \textit{NGC4593} & $6.86\pm0.21$ & $3.92\pm0.34$ & $4.25\pm0.60$ & $5.81\pm0.11$ & $-0.17\pm0.04$ & $-0.14\pm0.06$ & 1.08 & 3\\ 
    NGC4748 & $6.41\pm0.11$ & $3.36\pm0.58$ & $12.2\pm0.82$ & $0.97\pm0.05$ & $0.54\pm0.08$ & $1.10\pm0.04$ & 3.65 & <3\\
    NGC5548 & $7.72\pm0.02$ & $3.03\pm0.42$ & $2.67\pm0.42$ & $1.63\pm0.20$ & $0.27\pm0.08$ & $0.21\pm0.09$ & 0.88 & 3.5\\
    NGC6814 & $7.04\pm0.06$ & $4.97\pm0.35$ & $5.73\pm0.51$ & $1.19\pm0.35$ & $0.62\pm0.13$ & $0.68\pm0.13$ & 1.15 & 3\\
    NGC7469 & $6.96\pm0.05$ & $14.4\pm0.59$ & - & $2.07\pm0.14$ & $0.84\pm0.03$ & - & - & -\\
    PG1126-041 & $8.08\pm0.03$ & $4.00\pm0.21$ & $6.28\pm0.73$ & $1.33\pm0.07$ & $0.48\pm0.03$ & $0.67\pm0.06$ & 1.57 & <3\\
    PG1448+273 & $6.97\pm0.08$ & $2.55\pm0.15$ & $3.41\pm0.23$ & $0.24\pm0.01$ & $1.03\pm0.03$ & $1.15\pm0.03$ & 1.33 & <3\\
    \hline
    \end{tabular}
    \parbox[t]{\textwidth}{\textit{Notes:} (1) source name, (2) SMBH mass in units of $M_\odot$, (3), (4) and (5) are the $[\rm \ion{Ne}{v}]\,14\mu m$, $[\rm \ion{Ne}{v}]\,24\mu m$ and the Br$\gamma_{\rm broad}$ line fluxes in units of $10^{-17}\,{\rm W\,m^{-2}}$, (6) and (7) are the log of the $[\rm \ion{Ne}{v}]\,14\mu m$ and Br$\gamma_{\rm broad}$ and $[\rm \ion{Ne}{v}]\,24\mu m$ and Br$\gamma_{\rm broad}$ line ratios, (8) is the $[\rm \ion{Ne}{v}]\,24\mu m$ to $[\rm \ion{Ne}{v}]\,14\mu m$ ratio with (9) being the inferred electron density for [\ion{Ne}{v}] from the line ratio. The two sources where we used literature values for the $[\rm \ion{Ne}{v}]\,14\mu m$ line fluxes are identified with $^*$. The six sources where we took Br$\gamma_{\rm broad}$ line fluxes from \cite{Lamperti2017_BAT_BrG} are shown in italics.}
    \label{tab:sample}
\end{table*}

\section{Results}\label{Results}
We have studied the relation between the SMBH mass and the $\rm [\ion{Ne}{v}]14\mu m/Br\gamma_{\rm broad}$ and $\rm [\ion{Ne}{v}]24\mu m/Br\gamma_{\rm broad}$ line ratios in the $10^6-10^{8.5}\,{\rm M_\odot}$ SMBH mass range to investigate the use of CL as a SMBH mass proxy. All the lines are in the infrared, so that the effect of reddening and extinction is minimised. The broad line was chosen over the narrow line for the reasons discussed in \cite{Prieto2022_SiVI_BH}. In that work they found that when using the narrow line in the line ratio, the correlation with the SMBH mass disappears. They believe this is due to the larger emitting volume of the narrow \ion{H}{i} lines compared to the CL emitting region, which is believed to be in the inner parsecs of the central engine at the boundary of the BLR \citep[e.g.][]{Gravity_2021_CL_Region}.\par
\begin{figure*}
    \centering
    \includegraphics[width=0.4\textwidth]{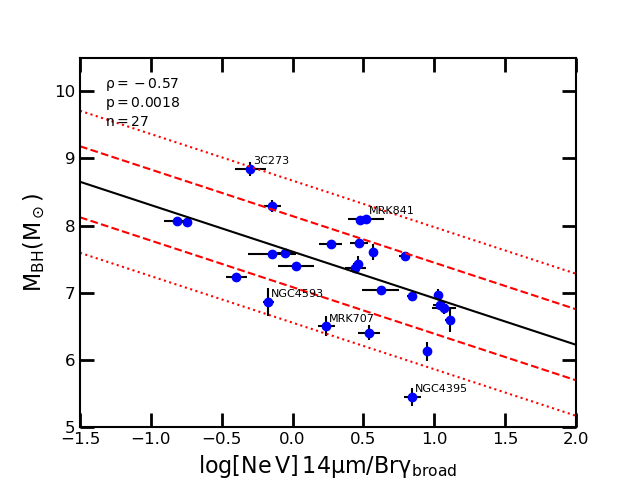}
    \includegraphics[width=0.4\textwidth]{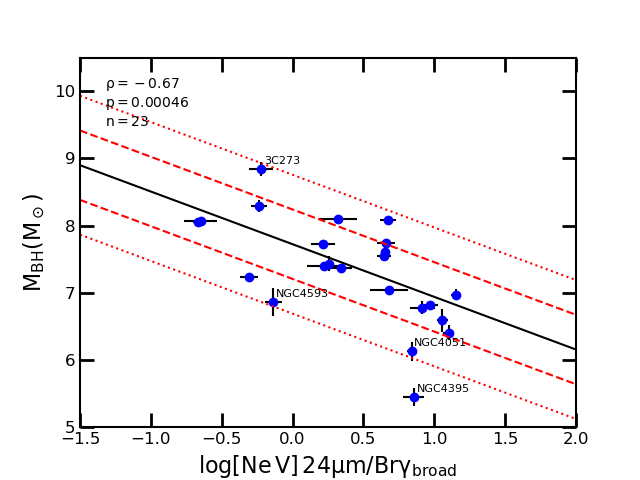}
    \caption{[\ion{Ne}{v}]$14\mu$m/Br$\gamma_{\rm broad}$ (left) and [\ion{Ne}{v}]$24\mu$m/Br$\gamma_{\rm broad}$ (right) versus black hole mass for the sources in our sample. The black line the best linear-fit to the data, and the red dashed and dotted lines represent the 1 and 2 $\sigma$ deviations, respectively.}
    \label{NeV_BH_Mass}
\end{figure*}

As shown in the left panel of Figure \ref{NeV_BH_Mass}, we find a correlation with the $\rm [\ion{Ne}{v}]14\mu m/Br\gamma_{\rm broad}$ line ratio over three orders of magnitude in BH mass. We perform a linear regression using the \textsc{LtsFit} package\footnote{\url{http://www-astro.physics.ox.ac.uk/~mxc/software/\#lts}} \citep{Cappellari2013_LTS_FIT}, which is a least squares fitting algorithm that allows for uncertainties in both axes and allows for outlier rejection. In our analysis only NGC4395 was excluded in both fits. From this fitting we find:\\
\begin{equation}
    {\rm log}M_{\rm BH}=(7.29\pm0.11)-(0.69\pm0.19)\times{\rm log}\left(\frac{[\ion{Ne}{v}]14\mu {\rm m}}{\rm Br\gamma_{\rm broad}}-0.47\right)
\end{equation}
with an observed scatter of 0.54 dex and an intrinsic scatter of $0.53\pm0.10$ dex. We performed a Spearman rank analysis and found a coefficient of -0.57 and a p-value of 0.0018, which indicates a moderate-to-strong anti-correlation which is unlikely to arise from random ordering of the data.\par
As shown in the right panel of Figure \ref{NeV_BH_Mass}, we also find a similar correlation with the $\rm [\ion{Ne}{v}]24\mu m/Br\gamma_{\rm broad}$ line ratio. Again, we perform a linear regression and find the following:
\begin{equation}
    {\rm log}M_{\rm BH}=(7.22\pm0.12)-(0.78\pm0.21)\times{\rm log}\left(\frac{[\ion{Ne}{v}]24\mu {\rm m}}{\rm Br\gamma_{\rm broad}}-0.64\right)
\end{equation}
with an observed scatter of 0.53 dex and an intrinsic scatter of $0.52\pm0.11$ dex. We performed a Spearman rank analysis and found a coefficient of -0.67 and a p-value of 0.0005, which indicates a moderate-to-strong anti-correlation which is unlikely to arise from random ordering of the data.
\par
Using the measured fluxes for the $[\rm \ion{Ne}{v}]\,14\mu m$ and the $[\rm \ion{Ne}{v}]\,24\mu m$ lines we can also measure the $[\rm \ion{Ne}{v}]\,24\mu m$ to $[\rm \ion{Ne}{v}]\,14\mu m$ ratio which can be used to infer the CL density. As shown in Table \ref{tab:sample}, we find flux ratios ranging from 0.64 to 3.65 with a median value of 1.08.
\par
The errors quoted in Table \ref{tab:sample} are only derived from reverberation mapping uncertainties which ignore systematic errors such as unknown BLR geometry, inclination, kinematics and the assumed radius-luminosity relation. To test whether these systematics could impact our relation, we assume an additional error of 0.5\,dex to account for these systematics \citep[e.g.][]{McGill2008_BH_Mass_Error,Shen2013_BH_Mass_Error,Pucha2025_BH_Mass_Error} and add this in quadrature to the existing BH mass error. We find that for both fits the fitting parameters remain the same with the same observed scatter of 0.54 and 0.53, respectively. The intrinsic scatter decreases to $0.21\pm0.20$ and $0.17\pm0.23$, respectively. This shows that even when adopting a more conservative BH mass error, the correlation remains.\par
We investigate if there are any correlations between the residuals of the fits and the AGN properties using a Spearman rank analysis. We look for correlations between the BH mass, the $\rm Br\gamma_{broad}$ luminosity and the CL ratio. The spearman rank analysis for the $\rm Br\gamma_{broad}$ luminosity have coefficients of -0.30 and -0.46 with p-values of 0.13 and 0.03 for the $[\rm \ion{Ne}{v}]\,14\mu m$ and the $[\rm \ion{Ne}{v}]\,24\mu m$ fits, respectively. This does suggest there is a weak anti-correlation with $\rm Br\gamma_{broad}$ luminosity for the $[\rm \ion{Ne}{v}]\,24\mu m$. However, this is likely driven by the one low luminosity point and the smaller sample than in the $[\rm \ion{Ne}{v}]\,14\mu m$ fit. The spearman rank analysis for the CL ratio gives coefficients of -0.17 and 0.00 with p-values of 0.13 and 1 for the $[\rm \ion{Ne}{v}]\,14\mu m$ and the $[\rm \ion{Ne}{v}]\,24\mu m$ fits respectively. This suggests there is no statistical correlation between the residuals of the fits and the CL ratio and thus the CL density. Finally, the spearman rank analysis of the BH mass gives coefficients of 0.79 and 0.79 and p-values of $9.9\times10^{-7}$ and $6.6\times10^{-6}$, respectively, for the $[\rm \ion{Ne}{v}]\,14\mu m$ and the $[\rm \ion{Ne}{v}]\,24\mu m$ fits. This suggests there is a strong correlation between the residuals of the fits and the BH masses.

\begin{figure}
    \centering
    \includegraphics[width=\linewidth]{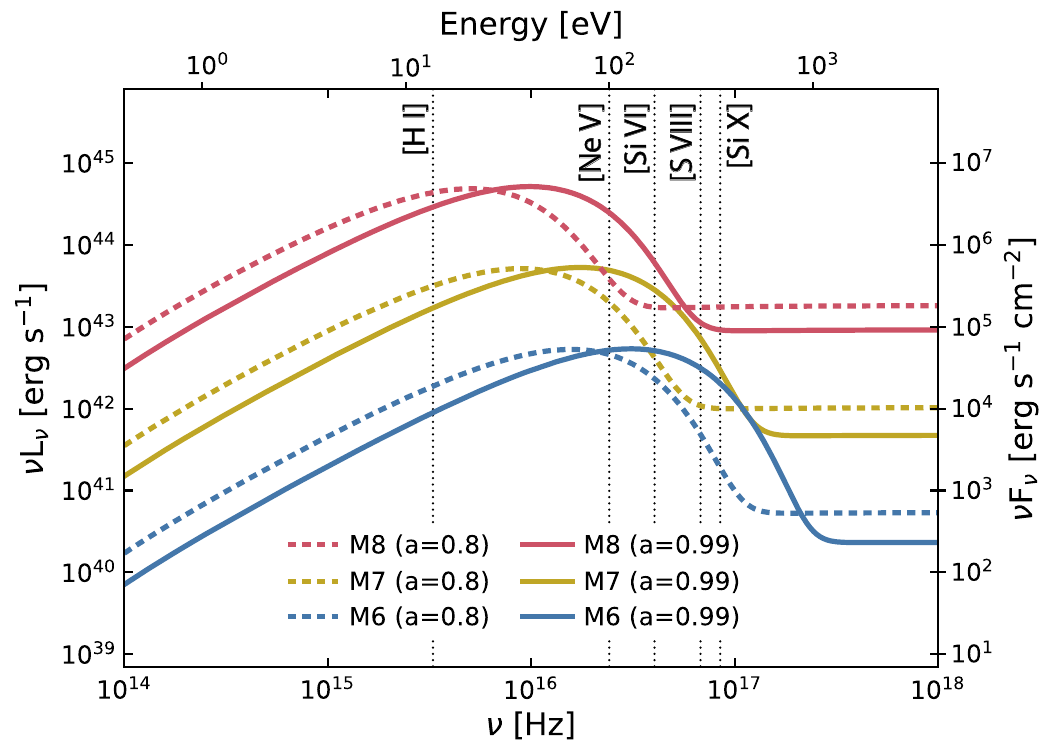}
    \caption{Generic ionising continuum following equation \ref{Continuum} for the range of BH masses studied in this work. Each curve corresponds to a specific BH mass ($10^6$ (purple), $10^7$ (green) and $10^8$ (red) $M_\odot$ and thus disc temperature with two spin values of 0.8 (dashed curve) and 0.99 (solid curves). Key coronal lines and their corresponding IPs are shown as vertical lines. Adapted from \cite{Prieto2022_SiVI_BH}.}
    \label{fig:Ionising_Cont}
\end{figure}

\section{Discussion}\label{Discussion}
\subsection{Physical interpretation}
As was presented in Section \ref{Results}, we found correlations between the $\rm [\ion{Ne}{v}]14\mu m/Br\gamma_{\rm broad}$ and $\rm [\ion{Ne}{v}]24\mu m/Br\gamma_{\rm broad}$ line ratios and the SMBH mass over three orders of magnitude. In \cite{Prieto2022_SiVI_BH}, a similar correlation was found in the same sample of sources using the [\ion{Si}{vi}]/Br$\gamma_{\rm broad}$ line ratio. The correlation they found is:
\begin{equation}
    {\rm log}M_{\rm BH}=(6.40\pm0.17)-(1.99\pm0.37)\times {\rm log}\left(\frac{[\ion{Si}{vi}]}{\rm Br\gamma_{\rm broad}}\right).
\end{equation}
They interpreted this relation as being driven by the temperature of the AGN accretion disc. The ionising continuum was parameterised with a Shakura-Sunyaev (SS) accretion disc \citep{ShakuraSunyaev1973_Accretion_Disc} and a power-law with a high and low energy cut-off to account for the rising of the continuum at high energies through the relation:
\begin{equation}\label{Continuum}
    F_\nu=\nu^{\alpha_{\rm uv}}\exp\left(\frac{-h\nu}{kT_{\rm disc}}\right)\exp\left(\frac{-kT_{\rm IR}}{h\nu}\right)+a\nu^{\alpha_{\rm x}}
\end{equation}

where the first term represents the SS accretion disc with a cutoff at $T_{\rm disc}$ and a power law with $\alpha_{\rm uv}=0.33$ to account for the low energy tail of the disc. The low energy limit of the disc is set by the IR-exponential with a cutoff at ${\rm k}T_{\rm IR}=0.01\,{\rm Ryd}$. The high-energy range is represented by a broken power law with a spectral index $\alpha_x=-1$ and a cutoff at 100\,keV. The parameter "a" controls the scaling between the SS disc relative to the high energy power law and is defined as the ratio of the $2$\,keV luminosity and the $2500\AA$ luminosity given by a power law with a spectral index $\alpha_{\rm OX}=-1.4$. The parameters above follow the generic AGN continuum used in CLOUDY C13.1 - Section 6.2 \cite{Ferland2017_CLOUDY}. CL are sensitive to different energy ranges in the ionising continuum. This is illustrated in Figure \ref{fig:Ionising_Cont}, which shows the parametrisation of the ionising continuum described above. Then assuming a thin accretion disc, the disc temperature, $T_{\rm disc}$ can be approximated as the following:
\begin{multline}\label{Disc_Temp}
    T_{\rm disc}=3.4\times10^5\,{\rm K}\left(\frac{M_{\rm BH}}{10^8\,M_\odot}\right)^{-1/4}\times\left(\frac{\left(\frac{dM}{dt}\right)}{0.1}\right)^{1/4}\times\left(\frac{\eta}{0.26}\right)^{-1/4}\\\times\left(\frac{Rin_G}{1.4}\right)^{-3/4}
\end{multline}
where $M_{\rm BH}$ is the BH mass, $dM/dt$ is the accretion rate in Eddington units, $\eta$ is the BH accretion efficiency and $Rin_G$ is the inner most stable radius in terms of the gravitational radius ($R_G$).\footnote{$R_G=G\,M_{\rm BH}/c^2$}. The disc temperature depends on the BH mass as shown in Equation \ref{Disc_Temp} and then the disc temperature sets the ionising continuum as shown in Equation \ref{Continuum} which in turn produces the CL emission. Therefore, in the BH mass range of $10^6-10^{8.5}\,\rm M_\odot$ the [\ion{Si}{vi}] line is produced optimally by the ionising continuum from the accretion disc as the number of photons with the required energy to produce [\ion{Si}{vi}] emission peaks in this range (shown in Figure \ref{fig:Ionising_Cont}). For this reason we expect a similar dependence with the $M_{\rm BH}$ when using the [\ion{Ne}{v}] coronal lines. 

As we are studying a similar sample as in \cite{Prieto2022_SiVI_BH} and the IP of the [\ion{Ne}{v}] lines is similar to that of the [\ion{Si}{vi}] line IP=97\,eV and IP=167\,eV, respectively, we believe that the disc temperature is also driving the relation we are seeing in our work where the [\ion{Ne}{v}] lines are also produced optimally by BH masses in the $10^6-10^{8.5}\,\rm M_\odot$ mass range. In this picture higher energy CLs like [\ion{S}{viii}] and [\ion{Si}{X}] (see Figure \ref{fig:Ionising_Cont}) will not correlate with BH mass in this range, as the disc temperature and thus the ionising continuum is not optimally producing these lines. In future work we will run photoionisation models that will further support this connection between the [\ion{Ne}{v}] coronal line production and the BH mass via the accretion disc temperature, allow for a comparison with the modeling of the [\ion{Si}{vi}] line done in \cite{Prieto2022_SiVI_BH} and to test which factors can impact this correlation (ionizing photon rate, gas density, distance of the ionised clouds etc). However, this is beyond the scope of the present paper.\par
We note that the findings in this work of a correlation between the [\ion{Ne}{v}] CL and the BH mass does differ from the findings in \cite{Bierschenk2024_CL_BASS}. In their work they studied several mid-IR CL including the $[\rm \ion{Ne}{v}]\,14\mu m$ and the $[\rm \ion{Ne}{v}]\,24\mu m$ lines in the BASS sample of AGN. They normalised the flux/luminosity of the [\ion{Ne}{v}] lines using both the $F_{\rm 14-150\,keV}$ and $L_{\rm bol}$ and looked for correlations with the AGN properties including the BH mass over a similar mass range ($10^{6.5}-10^{8.5}\,M_\odot$). They found no significant correlation between the [\ion{Ne}{v}] lines and the BH mass using either normalisation method though they did see a possible decrease of the [\ion{Ne}{v}] and bolometric luminosity ratio above $\sim10^{7.5}\,\rm M_\odot$. We believe that the lack of correlation in \cite{Bierschenk2024_CL_BASS} when confronted with our detected correlation is because the two studies probe different physics. In our work using the ratio of the $\rm [\ion{Ne}{v}]14\mu m/Br\gamma_{\rm broad}$ and $\rm [\ion{Ne}{v}]24\mu m/Br\gamma_{\rm broad}$ line ratios, we measure the shape of the ionising continuum (see Figure \ref{fig:Ionising_Cont}) which is influenced by the temperature of the accretion disc and thus the BH mass. On the other hand, in the work by \cite{Bierschenk2024_CL_BASS} when using the luminosity, they are measuring the integrated value of the ionising continuum and not directly probing the shape of the continuum.\par
There is some evidence that the production of CL could be tied to outflows \citep[e.g.][]{Matzko2025_CL_Outflows} and CL can be extended along the jet \citep[see][and references within]{Rodriguez-Adila2025_CL_Region}. If this is the case in our sources, this extended emission will be incorporated in our measured values due to the spatial resolution of \textit{Spitzer}. If outflows were the dominating production mechanism, then a correlation would not be expected given that the $\rm [\ion{Ne}{v}]14\mu m/Br\gamma_{\rm broad}$ and $\rm [\ion{Ne}{v}]24\mu m/Br\gamma_{\rm broad}$ line ratios would not be tied to the BH mass via the ionising continuum shape. However, the exact impact of outflows on the relations we present here needs to be investigated further.
\par
\subsection{Sources of scatter}
Using the measured $[\rm \ion{Ne}{v}]\,24\mu m$ to $[\rm \ion{Ne}{v}]\,14\mu m$ ratios in Table \ref{tab:sample} and using Figure 2 from \cite{Spinoglio2015_CL_Density} we are able to estimate the CL density. We find for most sources the value of $\frac{[\rm \ion{Ne}{v}]\,24\mu m}{[\rm \ion{Ne}{v}]\,14\mu m}$ implies a CL density in the order of $\rm n_e\sim10^3\,cm^{-3}$ with six sources (3C120, ARK120, NGC3516, NGC4748, PG1126-041, PG1448+273) with a high ratios $>1.25$ where the density could not be estimated and only an upper limit of $\lesssim10^3{\rm cm^{-3}}$ on the density could be placed. These density estimates are made assuming $\rm T=10,000\,K$. If a different temperature was assumed the density estimates would vary. These CL density estimates agree with those in previous work which also found a $\rm n_e\sim10^3\,cm^{-3}$ inferred from the $\frac{[\rm \ion{Ne}{v}]\,24\mu m}{[\rm \ion{Ne}{v}]\,14\mu m}$ flux ratio \citep[e.g.][]{Fernandez-Ontiveros2016_CL_Density}. In the sources that overlap between our sample and the sources studied in \cite{Fernandez-Ontiveros2016_CL_Density} the difference in the $\frac{[\rm \ion{Ne}{v}]\,24\mu m}{[\rm \ion{Ne}{v}]\,14\mu m}$ ratios varies from 6-60\% and agrees on the CL density estimates. We investigated whether the scatter in our relations is due to variations in the CL densities but found no clear evidence.\par
To study the impact of variability on our relation we collect fractional variability, $\rm F_{var}$\footnote{$F_{\rm var}=\sqrt{\frac{S^2-\langle\sigma_{\rm err}^2 \rangle}{\langle x \rangle ^2}}$ where $S^2$ is the variance of the measured fluxes, $\sigma_{\rm err}$ is the error on the fluxes and $\langle x \rangle$ is the mean of the measured fluxes.} , from $\rm H\beta$ reverberation mappings \citep{Santos-Lleo1997_RM,Collier1998_RM,Peterson1998_RM,Peterson2004_RM,Bentz2006_RM,Bentz2007_RM,Bentz2009_RM,Denney2009_RM,Denney2010_RM,Grier2012_RM,Du2014_RM,Kollatschny2014_RM,Peterson2014_RM,Hu2015_RM,Fausnaugh2017_RM,Park2017_RM,Bentz2021_RM} of our sources which we use as being representative of the BLR and thus the $\rm Br\gamma_{broad}$ variability. The maximum $\rm F_{var}$ found was 0.334 in NGC863 \citep{Peterson2004_RM}. We also estimate the [\ion{Ne}{v}] variability for both the $14\mu m$ and $24\mu m$ lines where multiple spectra were available from \textit{Spitzer}. We find for most sources if the [\ion{Ne}{v}] lines are variable, the fluxes are within the errors of the flux measurements. However, for one source (NGC4395) we do find variability in the $[\rm \ion{Ne}{v}\,14\mu m]$ line with an $\rm F_{var}$ values of 0.18. We also find variability in the $[\rm \ion{Ne}{v}\,24\mu m]$ line for three sources (MRK279, MRK335, NGC4395) with $\rm F_{var}$ values of 0.34, 0.27 and 0.21 respectively. It would be expected that if variability was observed in one of the [\ion{Ne}{v}] lines it would also be observed in the other line due to them arising from the same physical region though we do not see this. This could be explained if there are more noise in the region of the spectra where one of the lines is detected or one of the lines being poorly fitted. NGC4395 is the only source where we have a measurement of the $[\rm \ion{Ne}{v}\,14\mu m]$ and the $[\rm \ion{Ne}{v}\,24\mu m]$ variability and find they generally agree with each other. We choose to just use the $\rm H\beta$ as a proxy of the $\rm\rm Br\gamma_{broad}$ variability, which we propagate as an error term into the error in the $\rm Br\gamma_{broad}$ fluxes and then into the errors on the $\rm [\ion{Ne}{v}]14\mu m/Br\gamma_{\rm broad}$ and $\rm [\ion{Ne}{v}]24\mu m/Br\gamma_{\rm broad}$ line ratios. We find even in the most variable sources the errors still fall within the $2\sigma$ scatter of both relations indicating the variability is not the main driver of the scatter in our relations. To test whether the variability in NeV lines impacts our relations we include the NeV variability as another error term. We propagate this error into the error in $\rm [\ion{Ne}{v}]14\mu m/Br\gamma_{\rm broad}$ and $\rm [\ion{Ne}{v}]24\mu m/Br\gamma_{\rm broad}$ line ratios and again we find the errors fall within the $2\sigma$ scatter of both relations. This shows that even when we take into account the variability this does not produce outliers, and the relation still holds. In Figure \ref{NeV_BH_Mass} the error bars do not include this variability factor. 
\par
\subsection{Outliers}
As can been seen in both panels of Figure \ref{NeV_BH_Mass} one source (NGC4395) does not follow either relation and falls outside the scatter. To test whether this is truly where this sources lies in our relations or whether it a result of the quality of the \textit{Spitzer} observations we use the published JWST $[\rm \ion{Ne}{v}\,14\mu m]$ and $[\rm \ion{Ne}{v}\,24\mu m]$ from \cite{Goold2026_JWST_NGC4395} of NGC4395. We then recompute the $\rm [\ion{Ne}{v}]14\mu m/Br\gamma_{\rm broad}$ and $\rm [\ion{Ne}{v}]24\mu m/Br\gamma_{\rm broad}$ line ratios and find that the source barely deviates from the \textit{Spitzer} data. Moreover, it still falls outside the scatter in the relations. This indicates that our \textit{Spitzer} data is robust and this deviation is likely to be physical rather than an observational issue. NGC4395 is a dwarf galaxy and has the lowest BH mass in our sample of sources. This could explain why this source still follows the [\ion{Si}{vi}] relation but fails to follow the [\ion{Ne}{v}] relations due to slight difference in IP and can support the physical picture outlined in \cite{Prieto2022_SiVI_BH} where the production of CL lines is tied to the BH mass. Though it should be noted that for the [\ion{Si}{vi}] NGC4395 falls on the borderline of the 2$\sigma$ scatter. As [\ion{Si}{vi}] has a higher ionisation potential than the [\ion{Ne}{v}] lines, 167eV and 97eV respectively, the production of the [\ion{Si}{vi}] CL is more sensitive to lower BH masses. However, given the lower ionisation potential of the [\ion{Ne}{v}] lines they are less sensitive to the lower BH mass regime, and in this regime, the production of the [\ion{Ne}{v}] lines and the BH mass decouple explaining why NGC4395 does not follow our relations. This is supported by our study of the residuals in Section \ref{Results}, where we found a positive correlation between the residuals and the BH mass. However, the difference in IP between [\ion{Ne}{v}] and [\ion{Si}{vi}] is small and can be seen in Figure \ref{fig:Ionising_Cont} for a BH mass of $10^6\,M_\odot$ both lines sit near the peak of the ionising continuum. It would then be expected that both lines are produced optimally in NGC4395. It has been found that the narrow line region (NLR) of NGC4395 is very compact and likely dense \citep{Brum2019_NGC4395_NLR} and so could have very different physical conditions then the NLRs in the other sources studied in this work impacting the production of CL. Further investigation of the CL in NGC4395 are needed to understand why it does not follow the relations presented in this work.\par

\section{Conclusions}\label{Conclusion}
We have studied the [\ion{Ne}{v}] coronal lines of a sample of nearby AGN with bona fide BH masses from reveberation mapping and we have found a correlation between the $\rm [\ion{Ne}{v}]14\mu m/Br\gamma_{\rm broad}$ and $\rm [\ion{Ne}{v}]24\mu m/Br\gamma_{\rm broad}$ line ratios and the $\rm M_{BH}$ over three orders of magnitude  $\rm 10^6-10^8\,M_\odot$ (see Figure \ref{NeV_BH_Mass} and Section \ref{Results}). The relations we found have scatters of 0.54 and 0.53 dex, respectively which we believe is driven by differences in the accretion rate and spin in our sample of objects. This is compared to the scatter of the M-$\sigma$ relation which is 0.28-0.38\,dex depending on the sample and fitting method \citep[e.g.][]{KormendyHo2013_SMBH_Co_Evo,McConnelMa2013_M_Sigma,Saglie2016_M_Sigma}. These results support the interpretation that the $[\ion{Ne}{v}]14\mu m/Br\gamma_{\rm broad}$ and $\rm [\ion{Ne}{v}]24\mu m/Br\gamma_{\rm broad}$ relations are governed by the dependence of the accretion-disc temperature on the BH mass. This is consistent with the relation and photoionisation modeling presented in \cite{Prieto2022_SiVI_BH} and makes these relationships powerful tools to measure the SMBH masses in AGN where other methods are difficult to apply.\par
In the physical picture discussed in this work, the ionising continuum produced by the accretion disc depends on the BH mass and determines which CL can efficiently excited. As the BH mass increases to $10^8\,M_\odot$ the disc temperature is expected to be in the $10^5\,\rm K$ regime. This shifts the ionising continuum to lower energies. We therefore expected lower-ionisation CL to show correlations with BH mass in the higher mass regime. The BHs of this size are often found in ellipticals and bulge-dominated systems which often host Low Ionisation Nuclear Emitting Regions (LINER). However, coronal lines in these types of objects have been difficult to find \citep{MullerSanchez2013_LINER_CL,Mazzalay2014_LINER_CL} which might be explained by the spectral energy distribution of these objects being limited to disc temperatures $<10^5\,\rm K$ \citep{Fernandez-Ontiveros2023_LLAGN}. At the other end, higher ionisation lines would be expected to correlate with BH masses smaller then those studied in this work. The detections of the [\ion{Fe}{x}] coronal line with an IP=240\,eV in a large sample of dwarf galaxies by \cite{Molina2021_Dwarf_CL} provides a promising avenue to test this theory.\par
The relations shown in this work can only be reliably applied to Type 1 AGN including narrow line Type 1. This is due to reverberation being used to measure the BH masses in AGN and the need to normalise the coronal lines with the broad \ion{H}{i} emission. The finding in this work that the [\ion{Ne}{v}] can be used as a BH mass proxy could allow this to be extended to Type 2 AGN if an alternative normalisation can be found or a new method to estimate the broad line fluxes in obscured sources. Additionally, due to the [\ion{Ne}{v}] lines and the $\rm Br\gamma_{Broad}$ line being in different parts of the spectrum, it means that these lines cannot be measured simultaneously, which can induce additional systematic uncertainties. However, alternative lines, such as the Pfund lines, are often fainter than the $\rm Br\gamma_{broad}$ or fall outside the \textit{Spitzer} and JWST MIRI/MRS spectral range such as $\rm Br\alpha$. Finally, with JWST coming online and large-scale IR surveys taking place, it will allow a large number of SMBHs in AGN and out to higher redshift to be measured, furthering our understanding of the SMBH-host galaxy co-evolution and the role AGN play in this picture.   

\begin{acknowledgements}
JSE and AP acknowledges financial support from the Agencia Estatal de Investigación (AEI) through the Severo Ochoa Centre of Excellence accreditation awarded to the Instituto de Astrofísica de Canarias, grant CEX2025-001609-S, funded by MICIU/AEI/10.13039/501100011033. ARA acknowledges partial support from Conselho Nacional de Desenvolvimento Científico e Tecnológico (CNPq) through grant 313739/2023-4. \par
This research made use of MATPLOTLIB (https://matplotlib.org/), an open source visualisation package \citep{Hunter:2007_Matplotlib}, NUMPY (https://numpy.org/), an open source numerical computation
library \citep{harris2020array_Numpy}, SCIPY (https://scipy.org/) an scientific analysis library \citep{2020SciPy-NMeth} and PANDAS (https://pandas.pydata.org/), a data manipulation soft-
ware library \citep{mckinney-proc-scipy-2010,reback2020pandas}. 
\end{acknowledgements}

\bibliographystyle{bibtex/aa.bst}
\bibliography{REF.bib}

\begin{appendix}
\onecolumn
\section{\textit{Spitzer} spectra}
In this section we present the \textit{Spitzer} spectra of the [\ion{Ne}{v}]$\rm 14\mu m$ and the [\ion{Ne}{v}]$\rm 24\mu m$ lines for all the observations of the sources studied in this work with the associated Gaussian fits.  
\begin{figure*}[!ht]

\centering
    \includegraphics[width=0.28\linewidth]{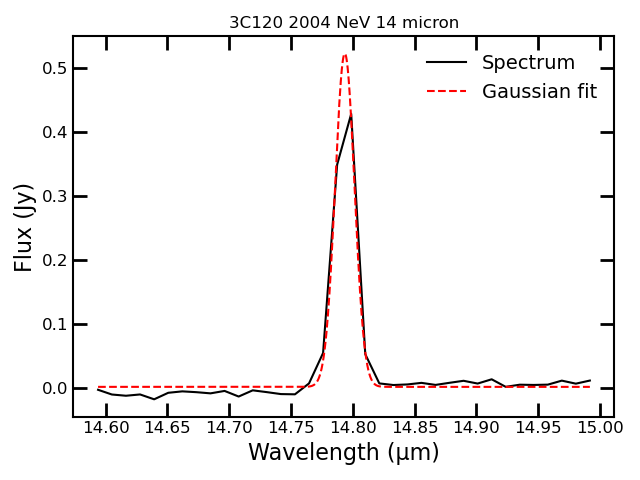}
    \includegraphics[width=0.28\linewidth]{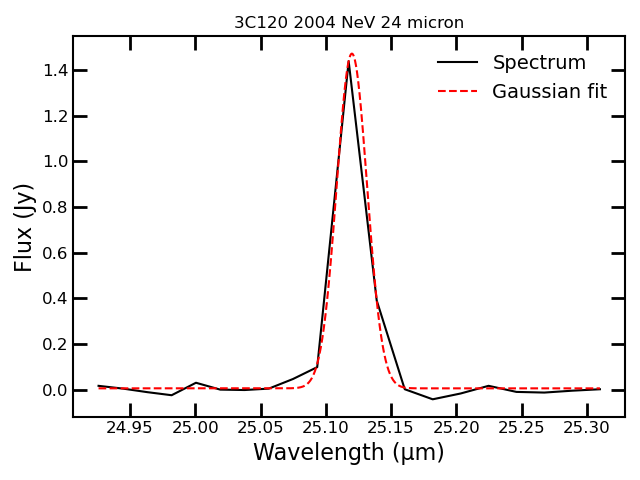}
    \includegraphics[width=0.28\linewidth]{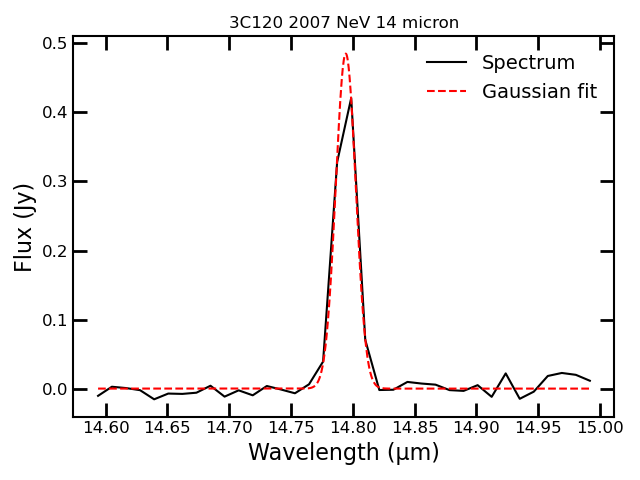}
    \includegraphics[width=0.28\linewidth]{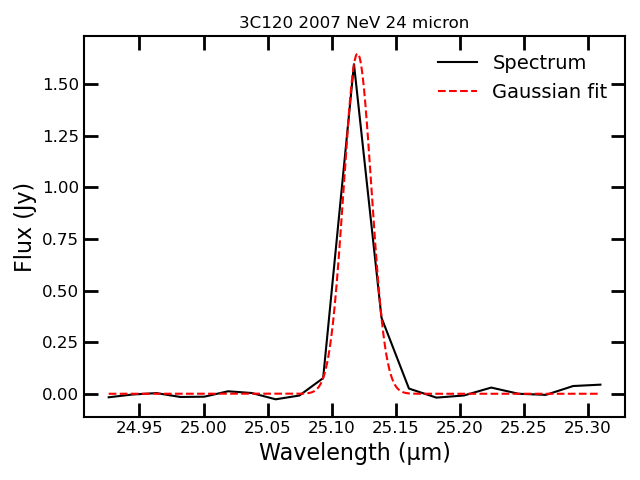}
    \includegraphics[width=0.28\linewidth]{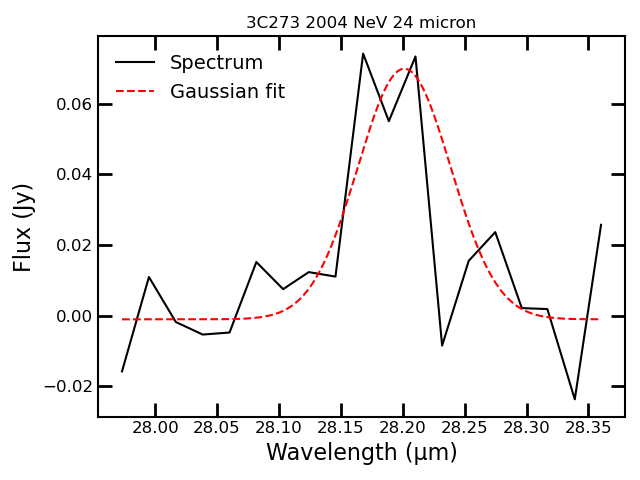}
    \includegraphics[width=0.28\linewidth]{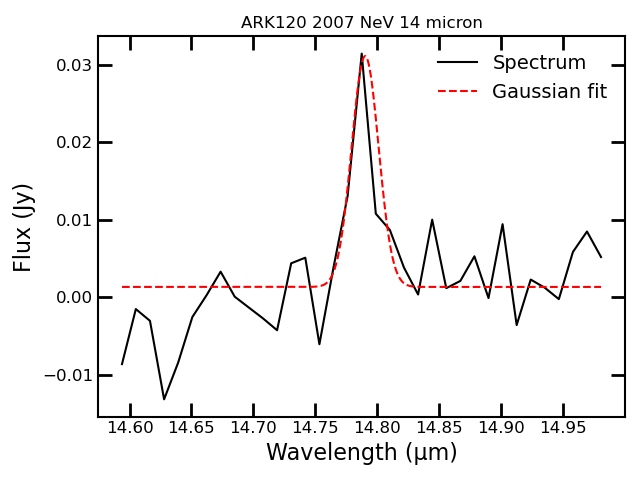}
    \includegraphics[width=0.28\linewidth]{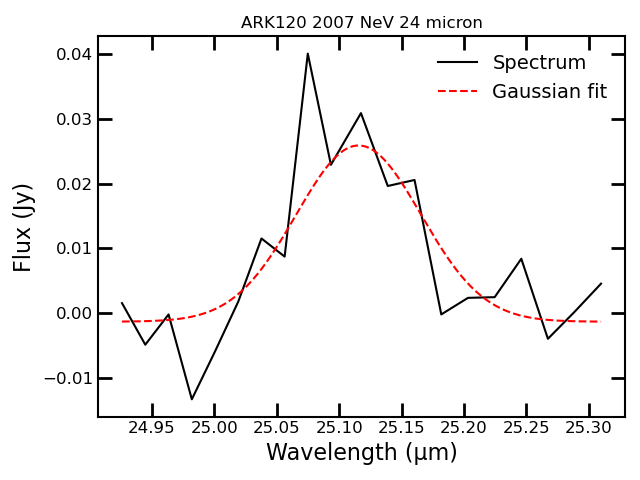}
    \includegraphics[width=0.28\linewidth]{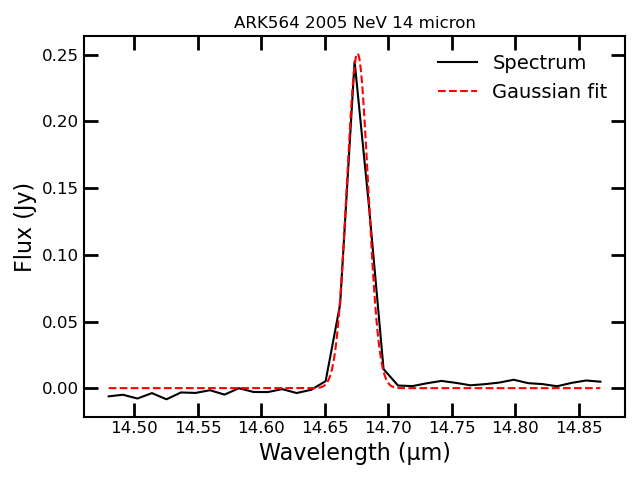}
    \includegraphics[width=0.28\linewidth]{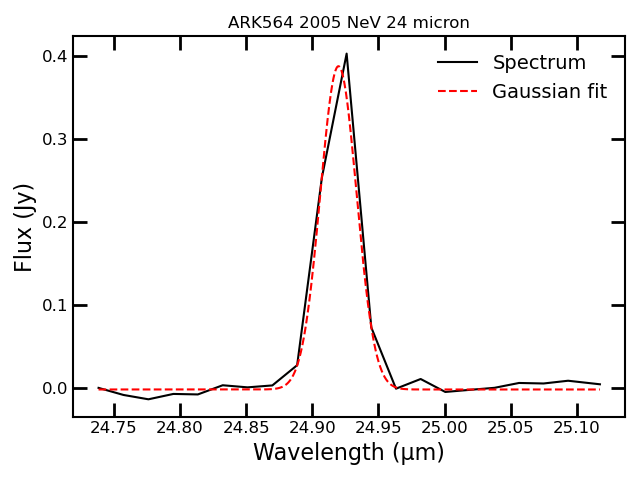}
    \includegraphics[width=0.28\linewidth]{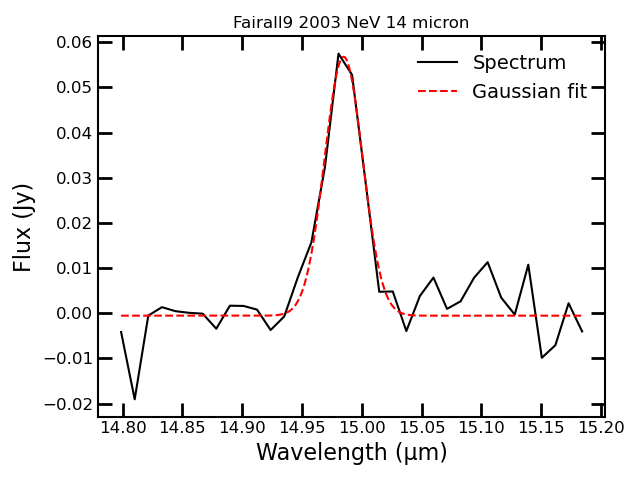}
    \includegraphics[width=0.28\linewidth]{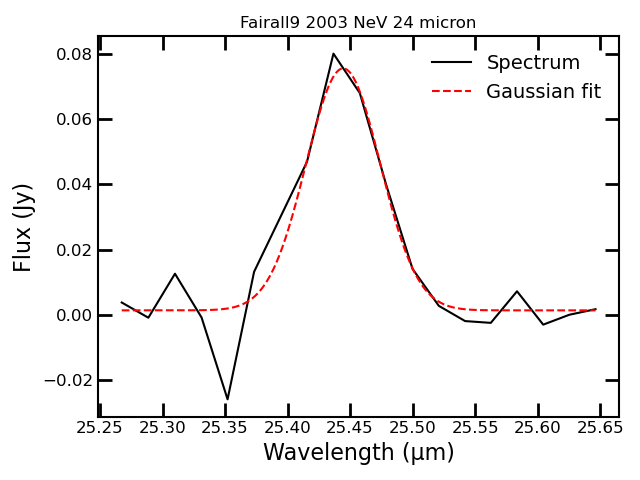}
    \includegraphics[width=0.28\linewidth]{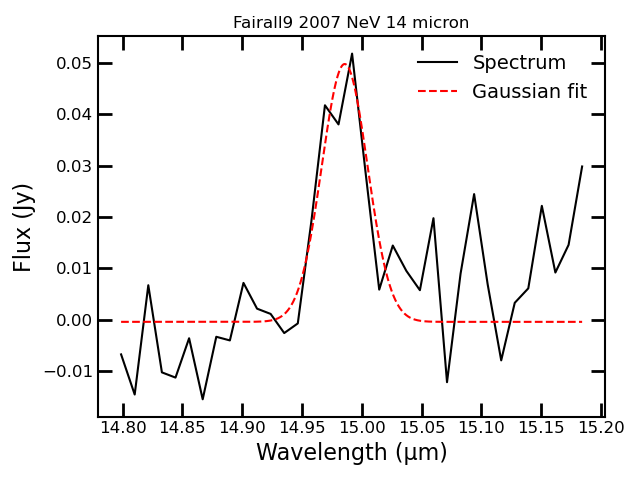}
    \includegraphics[width=0.28\linewidth]{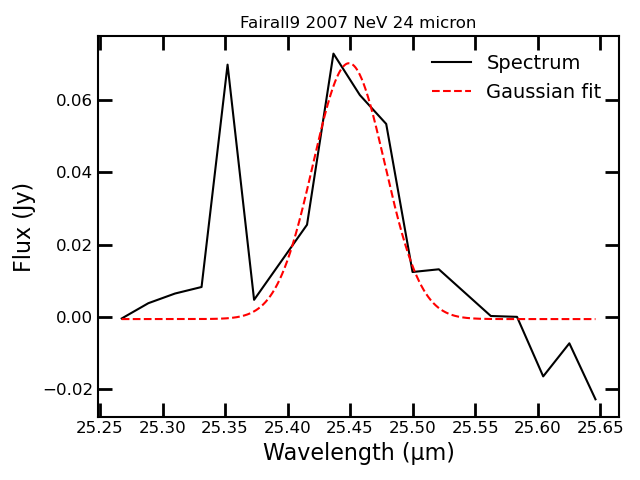}
    \includegraphics[width=0.28\linewidth]{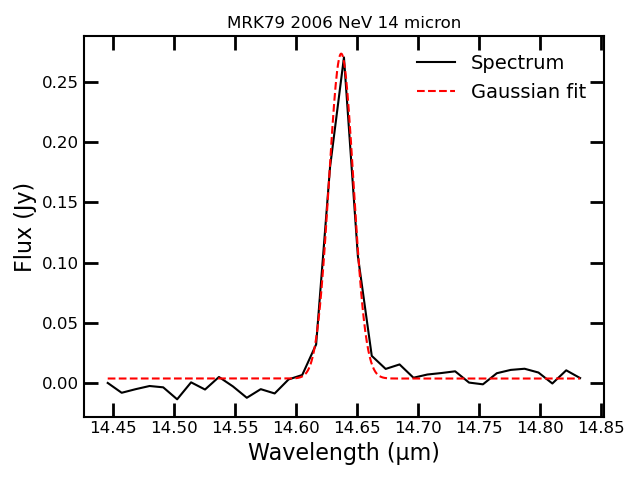}
    \includegraphics[width=0.28\linewidth]{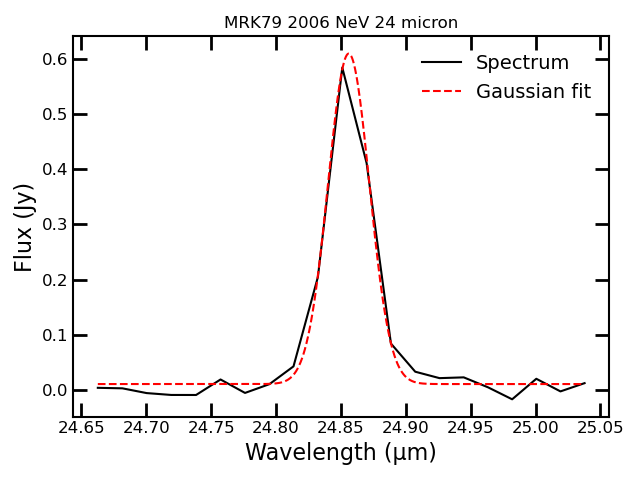}
    \includegraphics[width=0.28\linewidth]{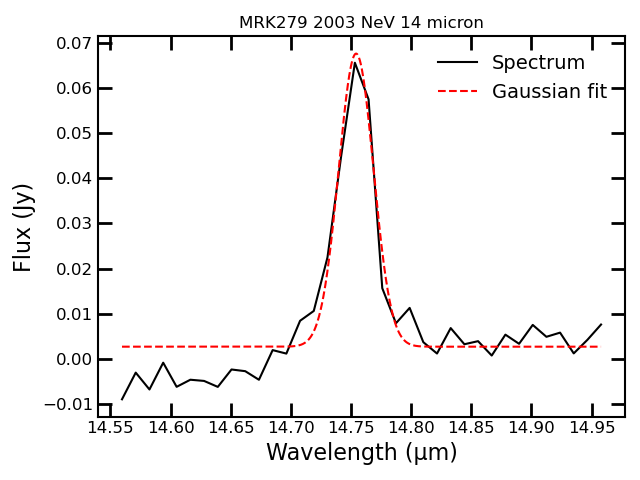}
    \includegraphics[width=0.28\linewidth]{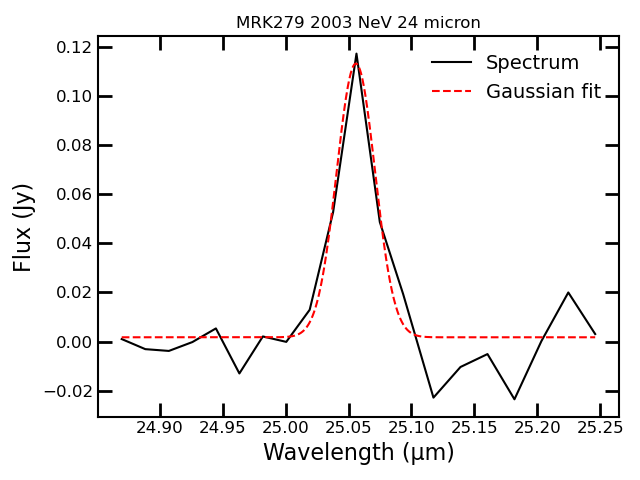}
    \includegraphics[width=0.28\linewidth]{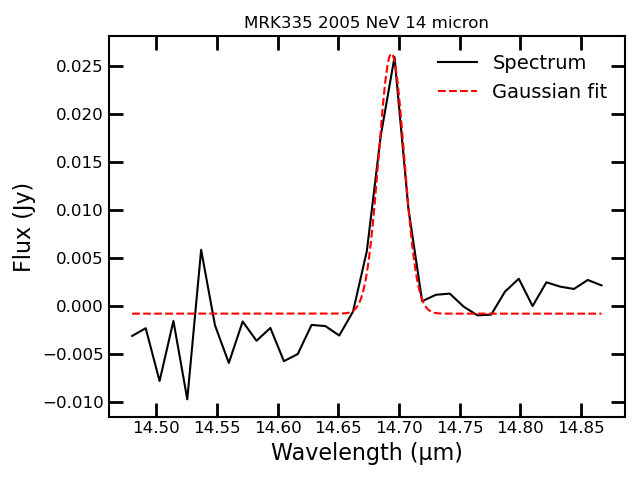}
\caption{Spitzer spectra of the [\ion{Ne}{v}]$\rm 14\mu m$ and the [\ion{Ne}{v}]$\rm 24\mu m$ line for all the sources studied here with the Gaussian fits shown as red dotted lines.}
\label{Spitzer_Fits}
\end{figure*}
\begin{figure*}
\centering
    \includegraphics[width=0.3\textwidth]{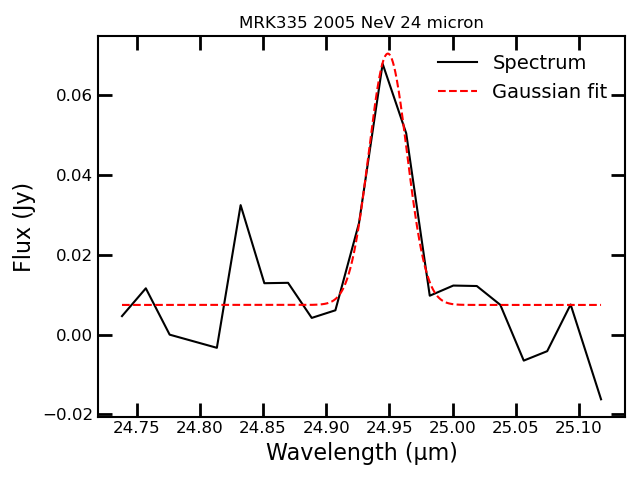}
    \includegraphics[width=0.3\textwidth]{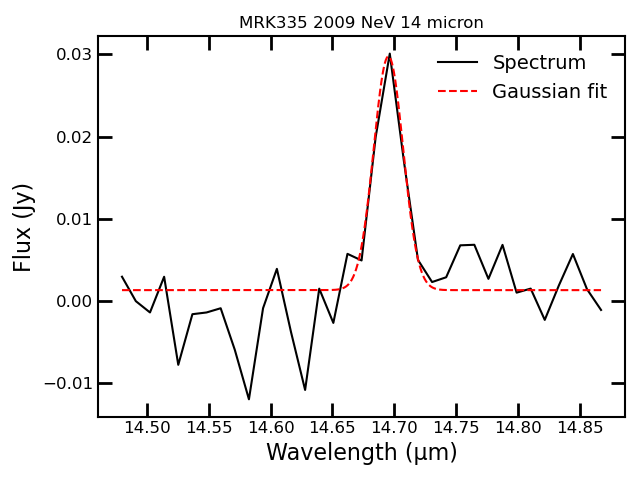}
    \includegraphics[width=0.3\textwidth]{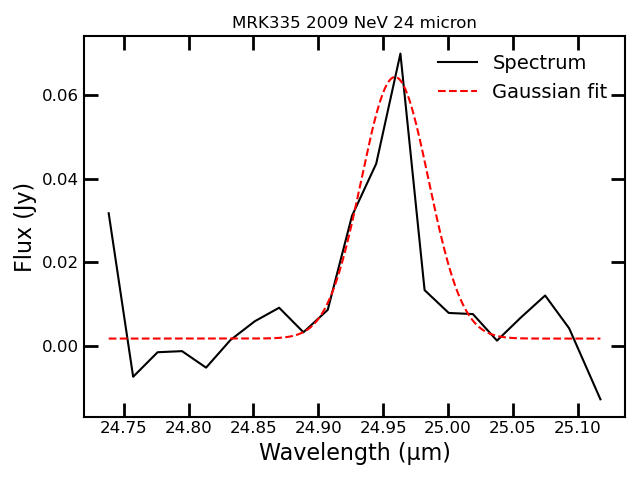}
    \includegraphics[width=0.3\textwidth]{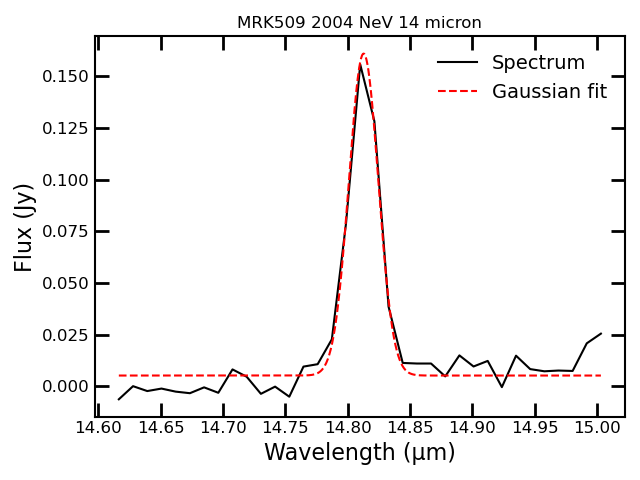}
    \includegraphics[width=0.3\textwidth]{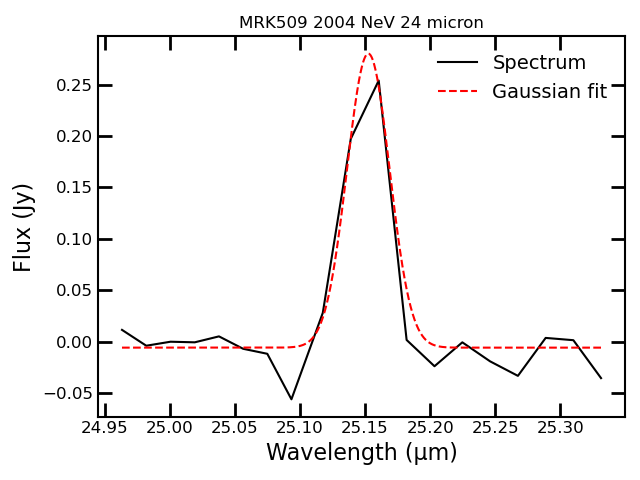}
    \includegraphics[width=0.3\textwidth]{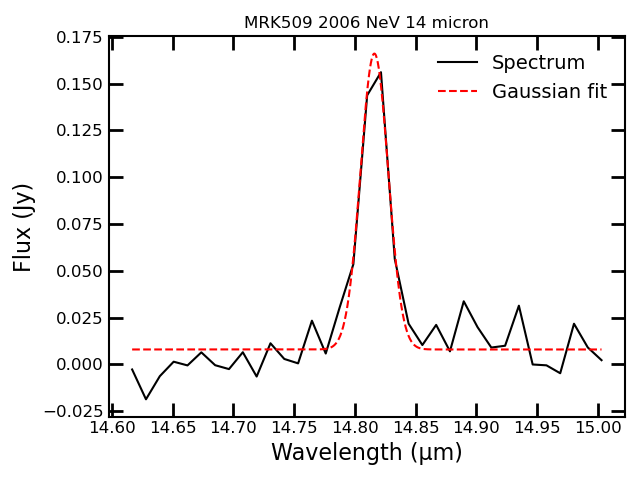}
    \includegraphics[width=0.3\textwidth]{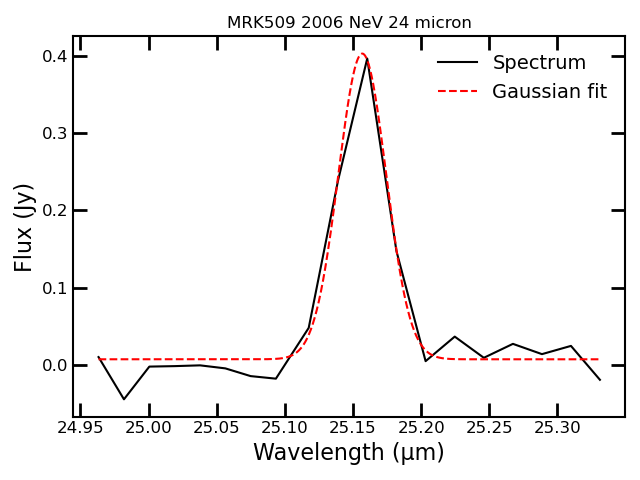}
    \includegraphics[width=0.3\textwidth]{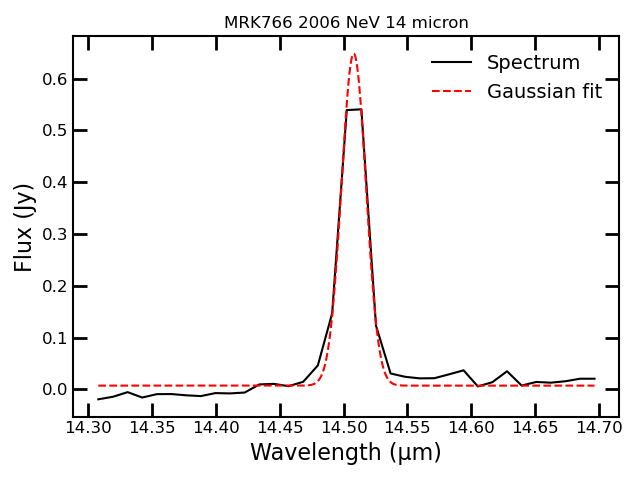}
    \includegraphics[width=0.3\textwidth]{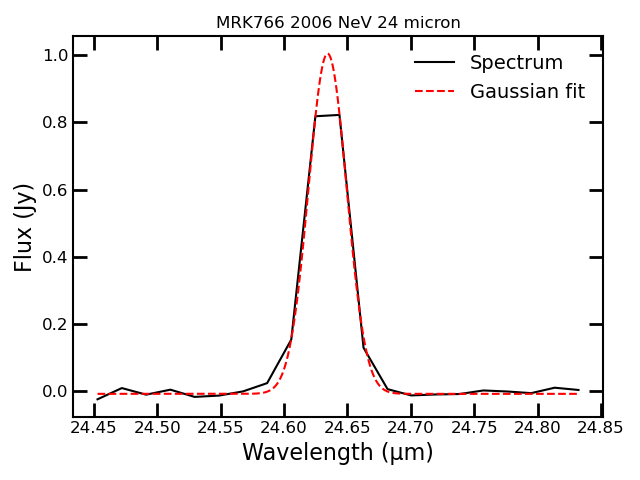}
    \includegraphics[width=0.3\textwidth]{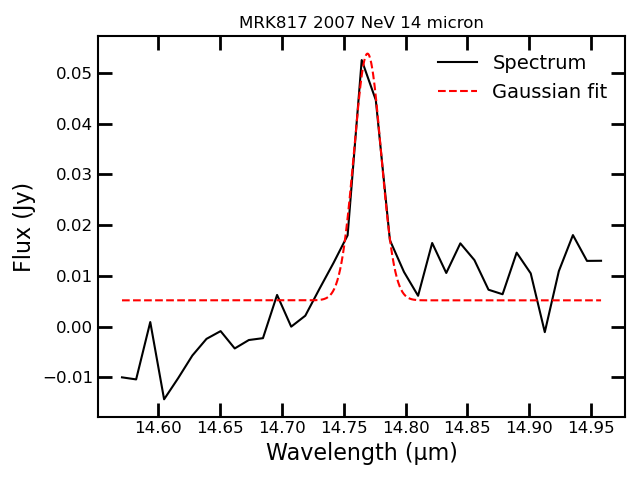}
    \includegraphics[width=0.3\textwidth]{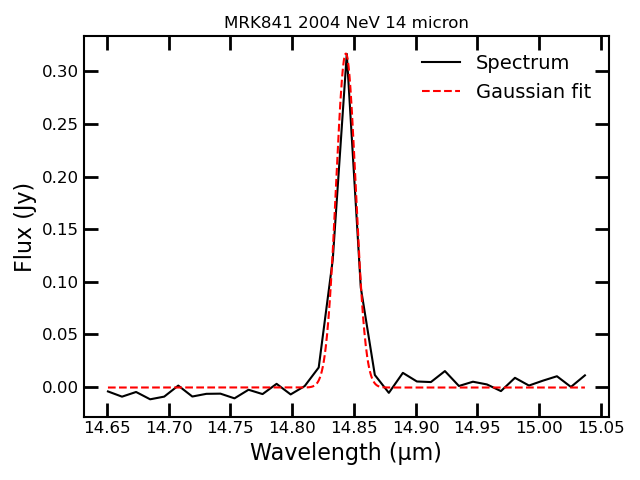}
    \includegraphics[width=0.3\textwidth]{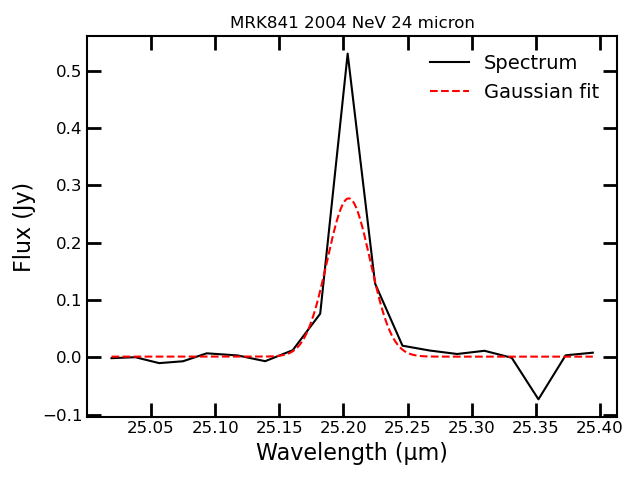}
    \includegraphics[width=0.3\textwidth]{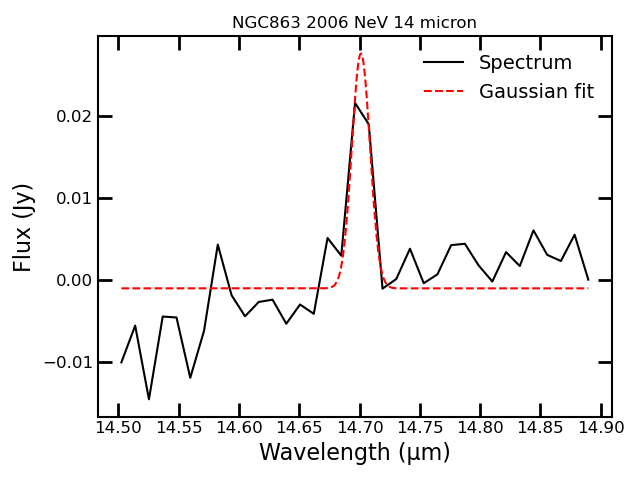}
    \includegraphics[width=0.3\textwidth]{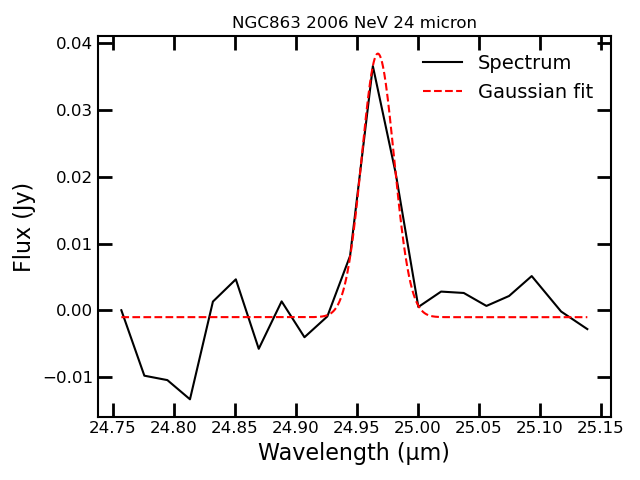}
    \includegraphics[width=0.3\textwidth]{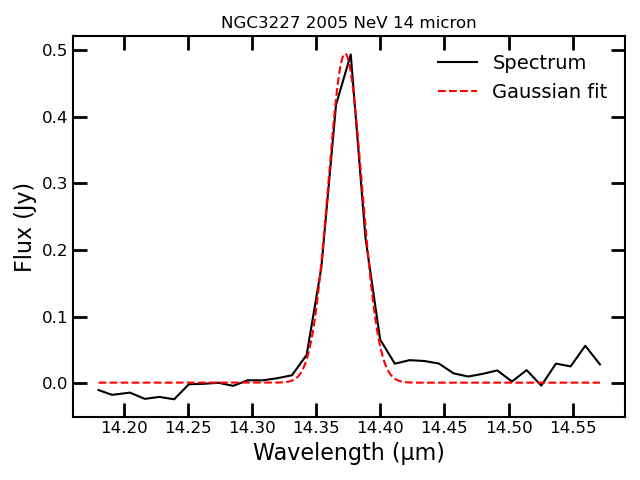}
    \includegraphics[width=0.3\textwidth]{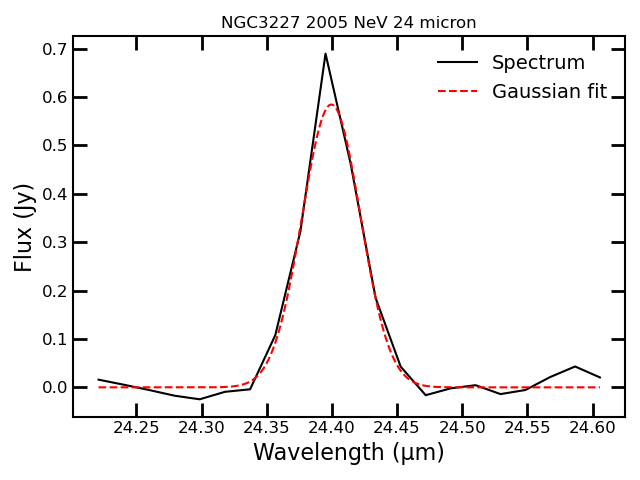}
    \includegraphics[width=0.3\textwidth]{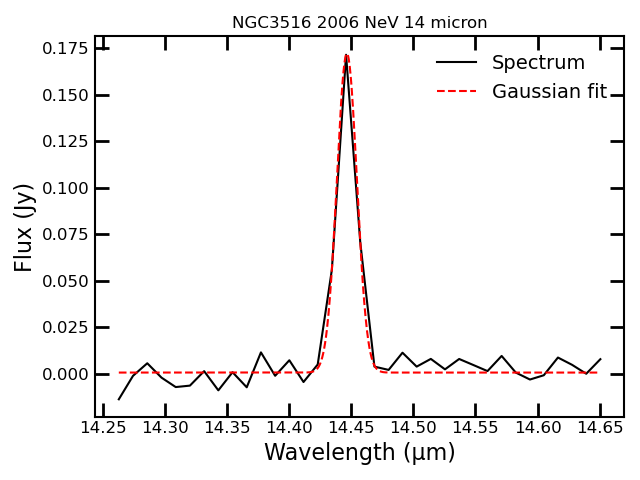}
    \includegraphics[width=0.3\textwidth]{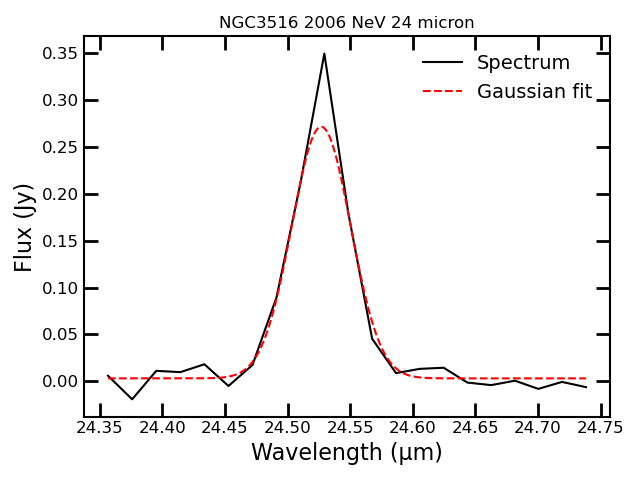}
    \caption*{Fig. A.1. continued}
    \end{figure*}
    \begin{figure*}
    \centering
    \includegraphics[width=0.3\textwidth]{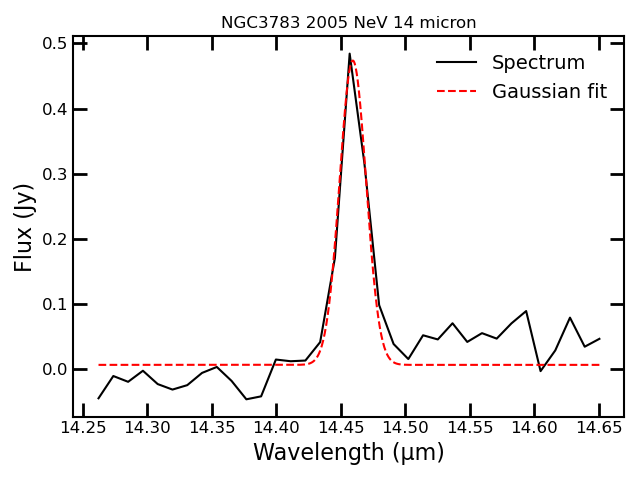}
    \includegraphics[width=0.3\textwidth]{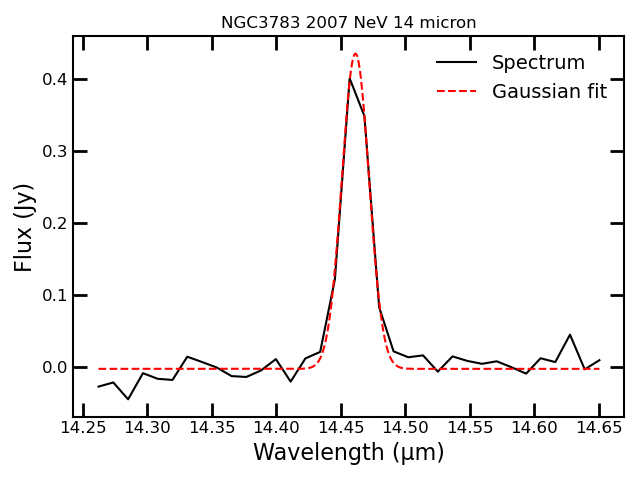}
    \includegraphics[width=0.3\textwidth]{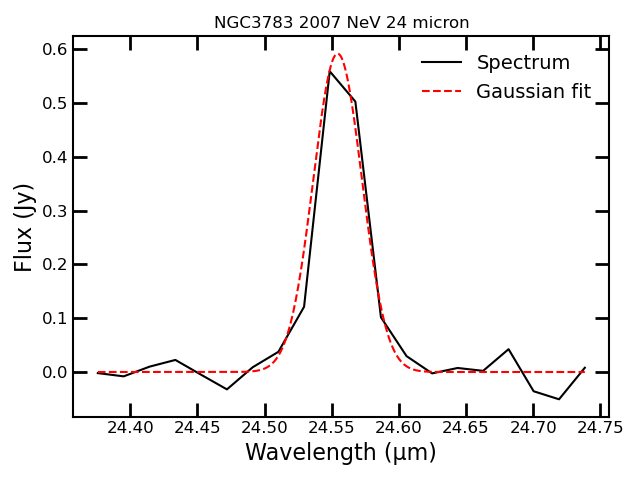}
    \includegraphics[width=0.3\textwidth]{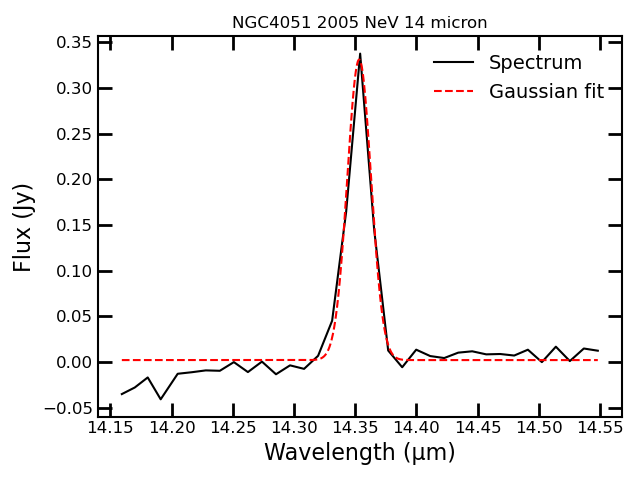}
    \includegraphics[width=0.3\textwidth]{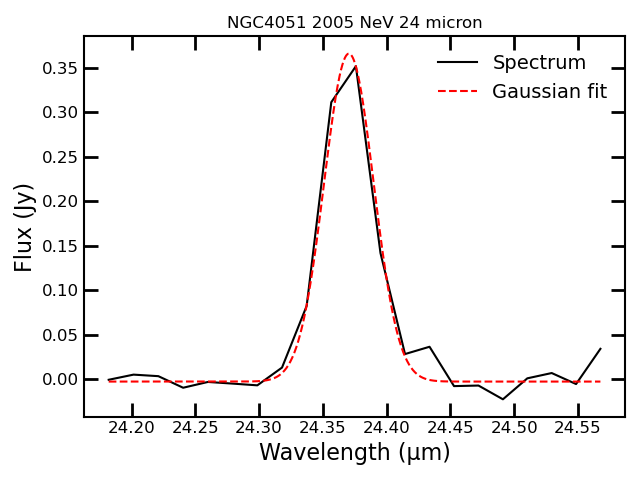}
    \includegraphics[width=0.3\textwidth]{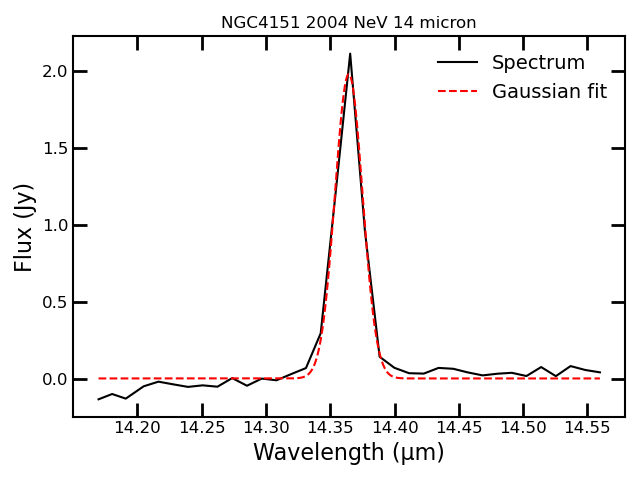}
    \includegraphics[width=0.3\textwidth]{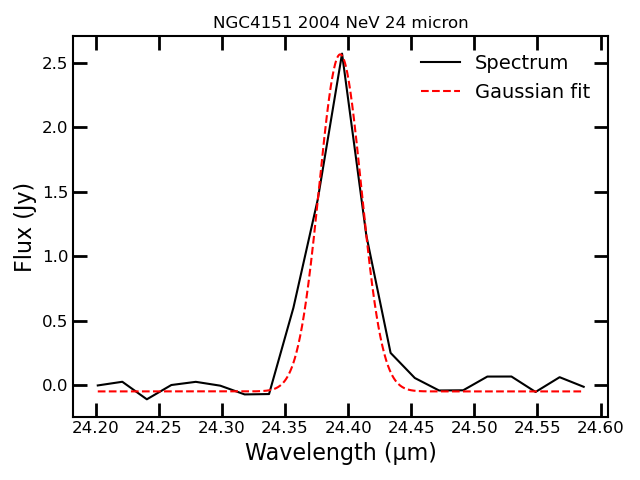}
    \includegraphics[width=0.3\textwidth]{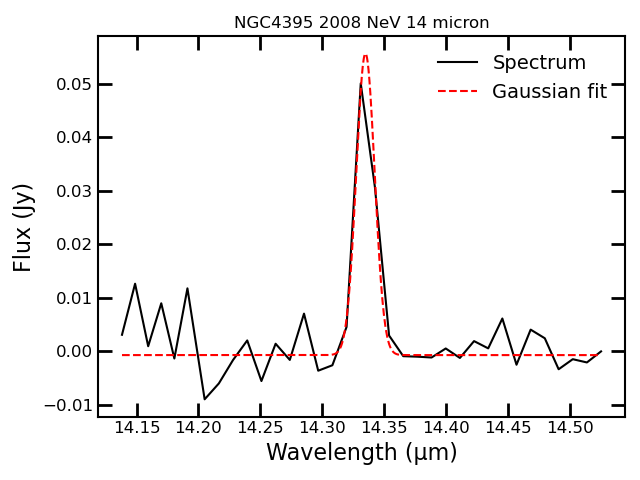}
    \includegraphics[width=0.3\textwidth]{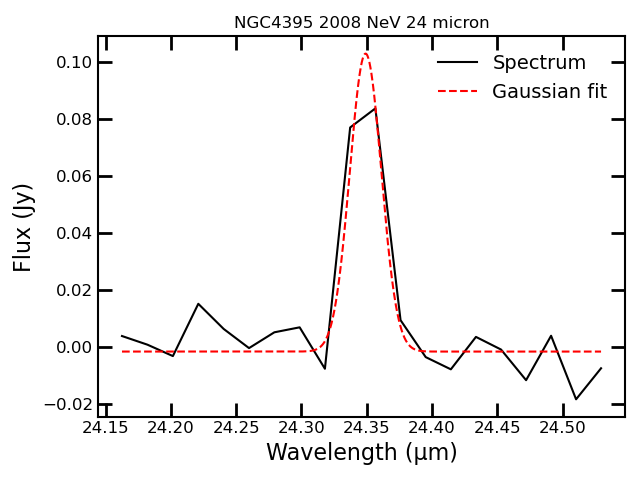}
    \includegraphics[width=0.3\textwidth]{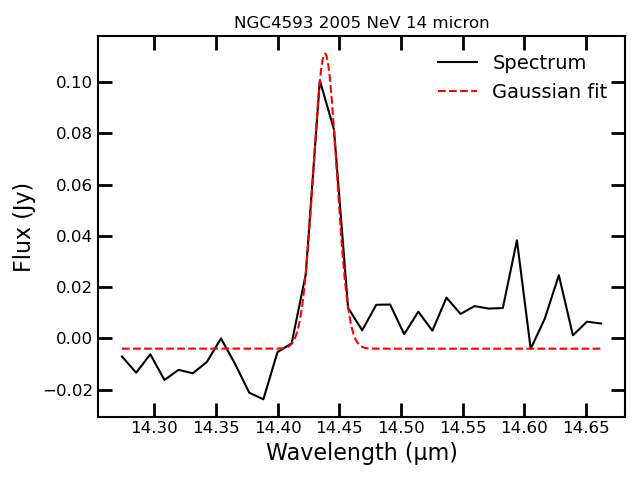}
    \includegraphics[width=0.3\textwidth]{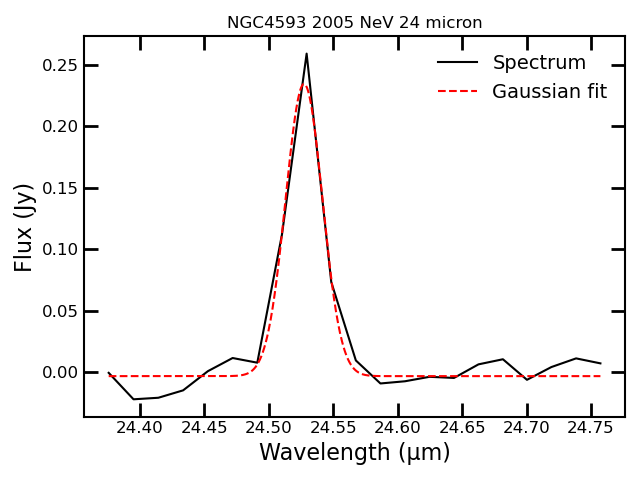}
    \includegraphics[width=0.3\textwidth]{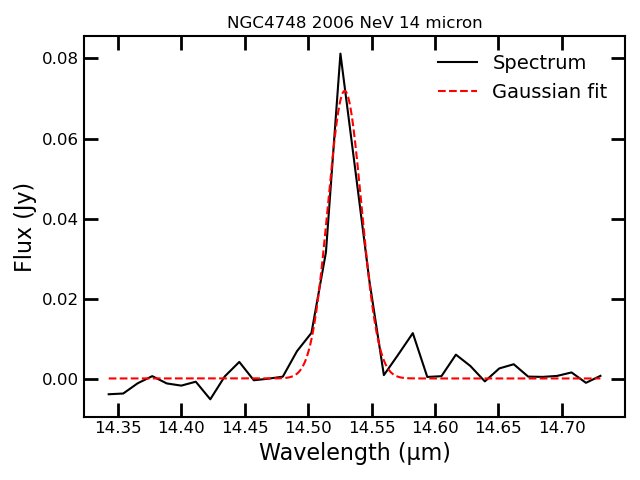}
    \includegraphics[width=0.3\textwidth]{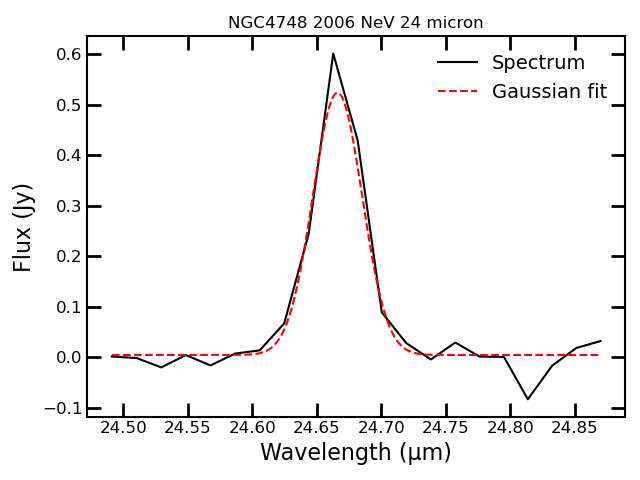}
    \includegraphics[width=0.3\textwidth]{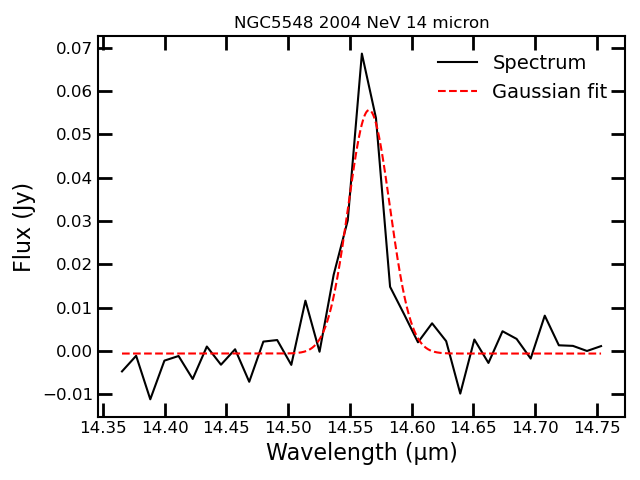}
    \includegraphics[width=0.3\textwidth]{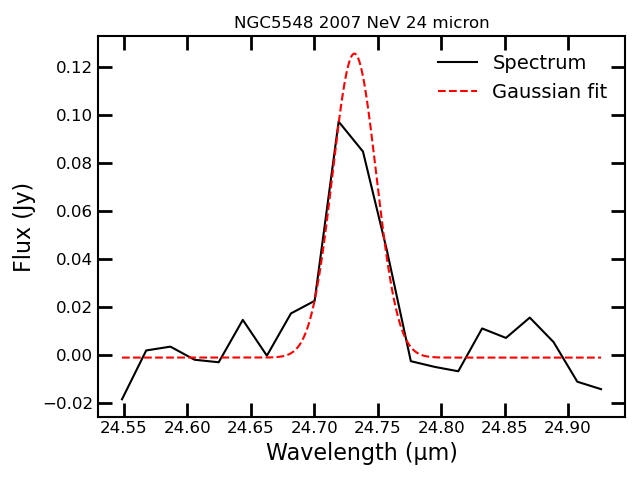}
    \includegraphics[width=0.3\textwidth]{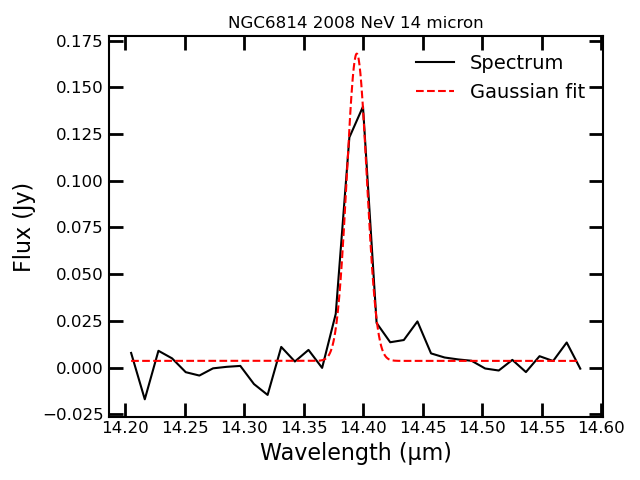}
    \includegraphics[width=0.3\textwidth]{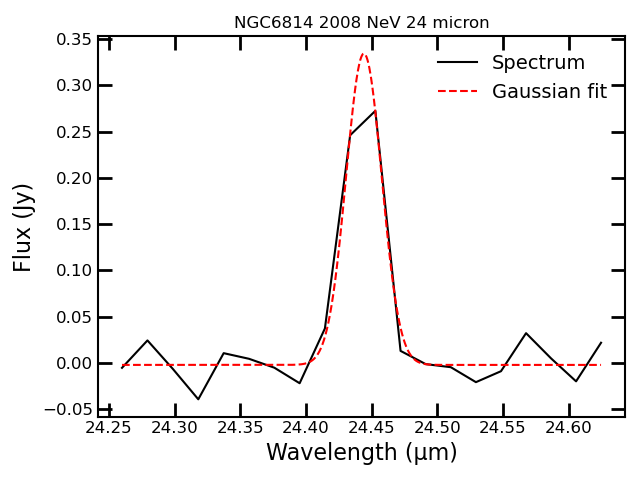}
    \includegraphics[width=0.3\textwidth]{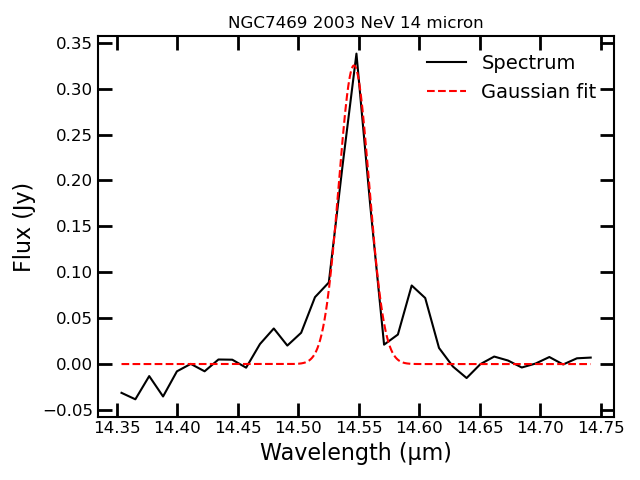}
    \caption*{Fig. A.1. continued}
    \end{figure*}
    \begin{figure*}
    \centering
    \includegraphics[width=0.3\textwidth]{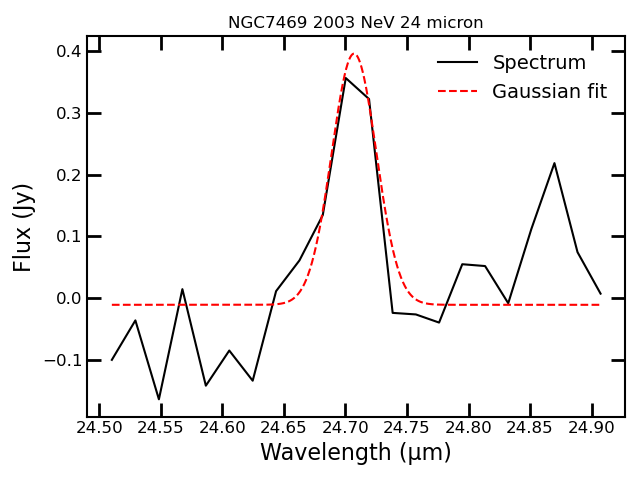}
    \includegraphics[width=0.3\textwidth]{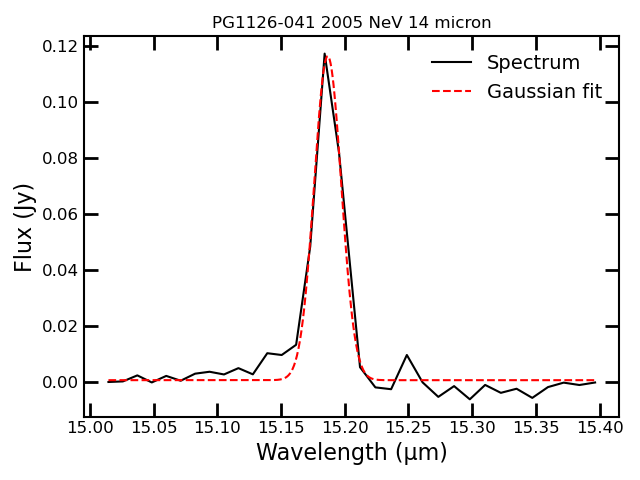}
    \includegraphics[width=0.3\textwidth]{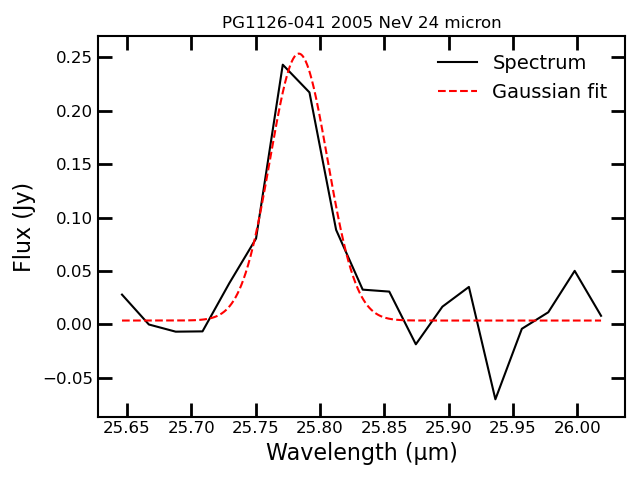}
    \includegraphics[width=0.3\textwidth]{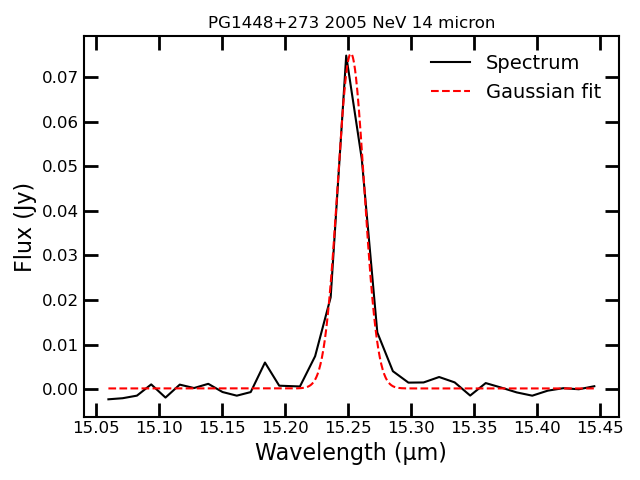}
    \includegraphics[width=0.3\textwidth]{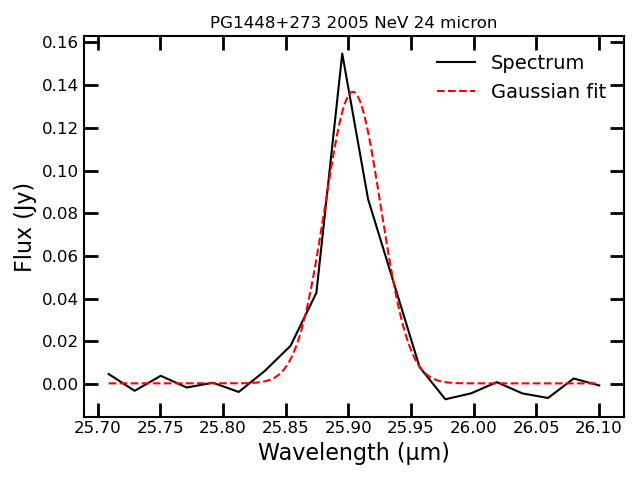}
    \caption*{Fig. A.1. continued}
\end{figure*}

\end{appendix}
\end{document}